\documentclass[aps,prb,groupedaddress,showpacs,twocolumn]{revtex4}
\usepackage{graphicx}
\usepackage{amsmath}
\usepackage{amssymb}
\usepackage{bbm}
\usepackage{xspace}
\usepackage{color}
\definecolor{orange}{RGB}{252,77,6}
\definecolor{brown}{RGB}{200,127,50}
\definecolor{green1}{RGB}{00,100,00}
\definecolor{green2}{RGB}{00,150,00}
\definecolor{green3}{RGB}{00,200,00}
\definecolor{green4}{RGB}{00,250,00}

\newcommand{\fig}[1]{Fig.\thinspace{}\ref{#1}}
\newcommand{\eq}[1]{eq.\thinspace{}(\ref{#1})}

\newcommand{\se}{Sec.\@\xspace}

\newcommand{\etal}[0]{\textit{et al.}}
\newcommand{\tcite}[1]{Ref.~\onlinecite{#1}}

\newcommand{\uu}{1\hspace{-3pt}1}

\newcommand{\nag}{{\phantom{\dag}}}

\begin{document}

%\preprint{}

\title{Effects of electronic correlations and magnetic field on a molecular ring out of equilibrium}


\author{Martin Nuss}
\email[]{martin.nuss@tugraz.at}
\affiliation{Institute of Theoretical and Computational Physics, Graz University of Technology, 8010 Graz, Austria}
\author{Wolfgang von der Linden}
\affiliation{Institute of Theoretical and Computational Physics, Graz University of Technology, 8010 Graz, Austria}
\author{Enrico Arrigoni}
\affiliation{Institute of Theoretical and Computational Physics, Graz University of Technology, 8010 Graz, Austria}
%\thanks{}


\date{\today}

\begin{abstract}
We study effects of electron-electron interactions on the steady-state characteristics of a hexagonal molecular ring in a magnetic field, as a model for a benzene molecular junction. The system is driven out of equilibrium by applying a bias voltage across two metallic leads. We employ a model Hamiltonian approach to evaluate the effects of on-site as well as nearest-neighbour density-density type interactions in a physically relevant parameter regime. Results for the steady-state current, charge density and magnetization in three different junction setups (para, meta and ortho) are presented. Our findings indicate that interactions beyond the mean-field level renormalize voltage thresholds as well as current plateaus. Electron-electron interactions lead to substantial charge redistribution as compared to the mean-field results. We identify a strong response of the circular current on the electronic structure of the metallic leads. Our results are obtained by steady-state Cluster Perturbation Theory, a 
systematically improvable approximation to study interacting molecular junctions out of equilibrium, even in magnetic fields. Within this framework general expressions for the current, charge density and magnetization in the steady-state are derived. The method is flexible and fast and can straight-forwardly be applied to effective models as obtained from {\em ab-initio} calculations. 
\end{abstract}

%71.27+a strongly correlated systems
%71.10.-w theories and models of condensed matter
%71.15.-m condensed matter calculation methods
%72.15.Qm Kondo Effect
%71.55.Ak local magnetic moments in metals
%73.63.Kv       Quantum dots
%73.23.-b       Electronic transport in mesoscopic systems
%72.10.Fk Scattering by point defects, dislocations, surfaces, and other imperfections (including Kondo effect)
%71.15.-m condensed matter calculation methods
%72.15.Qm Kondo Effect
%71.27+a strongly correlated systems
%Biomolecular electronics, 85.65.+h
%Electronic transport nanoscale materials, 73.63.-b
%Persistent currents (mesoscopic systems), 73.23.Ra
%Perturbation theory applied to electronic structure of solids, 71.15.-m
%71.10.-w theories and models of condensed matter
%\keywords{}

\maketitle

\section{Introduction}\label{sec:introduction}
% MINIATURIZATION TO MOLECULAR DEVICES STEP 1) BASIC UNDERSTANDING OF TRANSPORT
Miniaturization as a performance enhancing concept in micro-electronics may be advanced by the introduction of nanoscale molecular devices in what is today known as the concept of molecular-electronics.~\cite{cuniberti_2005} Recent years fostered fascinating advances in experimental control of fabrication,~\cite{smit_measurement_2002} assembling~\cite{park_coulomb_2002} as well as contacting~\cite{liang_kondo_2002,Agrait200381} on a molecular level. These achievements in combination with ever improving measurement techniques~\cite{venkataraman_dependence_2006} lead to a plethora of important insights into the basic mechanisms of electrical transport across molecular junctions.~\cite{Nitzan30052003} Understanding these transport characteristics is a major focus of today's experimental as well as theoretical ventures and establishes the very basis for possible future device engineering.

% PARTICULAR DEVICE - HOW IS IT REALIZED, WHAT IS MEASURED
A relevant and still relatively simple molecular junction is comprised of a ring-shaped molecule in between two metallic leads. For the particular molecule we have in mind the aromatic complex benzene (C$_6$H$_6$). Such setups have been realized using the mechanically controllable break junction technique (MCBJ)~\cite{Agrait200381} and are stable on time-scales required for transport experiments. Measurement of the transport characteristics has been achieved for benzene bound by thiol anchor groups to gold electrodes~\cite{Reed10101997,PhysRevLett.98.176807,Cuevas2010} as well as for benzene directly connected to platinum leads.~\cite{PhysRevLett.101.046801} These experiments typically grant access to the current voltage characteristics, conductance, higher derivatives of the current or shot noise, albeit often in a statistical way.~\cite{PhysRevLett.98.176807} It is likely that molecular ring junctions will find technological applications in the foreseeable future in the form of single electron 
transistors,~\cite{song_observation_2009,doi:10.1021/jz200535j} in quantum interference (QI) based electronics,~\cite{doi:10.1021/nl8016175} or as data storage devices~\cite{0957-4484-14-2-307}. Naturally such junctions are dominated by quantum mechanical effects in an out of equilibrium context which promotes predicting the outcome of experiments to highly non trivial theoretical task.

% PARTICULAR DEVICE - WHAT IS KNOWN FROM THEORY
For benzene based molecular junctions, general understanding of the noninteracting device is available in literature.~\cite{PhysRevB.85.155440,doi:10.1021/jz200862r} Electronic transport in $\pi$ conjugated systems is special due to QI effects.\cite{Solomon2008,doi:10.1021/nl101688a} Magnetic fields add to the rich Aharonov-Bohm physics~\cite{PhysRev.115.485,PhysRev.123.1511} of quantum ring structures by inducing for example persistent currents.~\cite{Viefers20041} Electronic correlations are important~\cite{doi:10.1021/nl201042m} due to the confined geometry and have been recently studied in equilibrium and linear response using the Hirsch-Fye quantum monte carlo (QMC) method~\cite{Valli2013} as well as dynamical mean field theory (DMFT).~\cite{PhysRevB.86.115418,Valli2013} While the basic features of QI effects can be understood from noninteracting calculations, the interplay of QI and electronic correlations can become non trivial.~\cite{PhysRevB.77.201406,PhysRevB.79.235404,PhysRevB.79.245125,doi:10.1021/nl901554s,ANDP:ANDP201100266}. The remarkable property of negative differential conductance has been reported and explained in devices considering electron-electron interactions~\cite{Leijnse_2011,PhysRevB.79.235404} using generalized master equation approaches. In addition Green's function techniques have been applied within various approximations, especially within a combination with {\em ab-initio} techniques.~\cite{PhysRevB.86.195425} In gated devices even non perturbative many-body features like the Kondo effect~\cite{hewson_kondo_1997} have been reported.~\cite{Bohr_2012,park_coulomb_2002,liang_kondo_2002,PhysRevLett.95.256803,Tosi2012}

% HOW THEORY CAN CONTRIBUTE - DFT+NEGF
Investigating transport characteristics for molecular junctions out of
equilibrium~\cite{Agrait200381,B922000C} is a very active field in
modern theoretical physics. Although much progress has been made
lately,~\cite{Datta_2005,Ventra_2008,Ferry_2009,Nazarov_2009}
reproducing experiments~\cite{doi:10.1021/nl052373+,Cuevas2010} in a
qualitative and quantitative way remains an elusive goal. In going
beyond semi-classical treatment~\cite{richter_semiclassical_1999} and
combining non equilibrium Green's function (NEGF)
methods~\cite{0022-3719-4-8-018} with density functional theory
(DFT)~\cite{PhysRev.136.B864,PhysRev.140.A1133} much effort is devoted
to the study of molecular junctions in a combined DFT+NEGF
scheme.~\cite{PhysRevB.75.205413,PhysRevB.63.245407,PhysRevB.65.165401,PhysRevB.67.195315,PhysRevB.73.085414,PhysRevB.69.035108,Thygesen2005111,0953-8984-18-4-019,doi:10.1021/jp8075854,Arnold_2007,footnote1}
Such {\em ab-initio} approaches can be considered today's gold
standard for weakly correlated molecular junction calculations. However, they tend to overestimate currents in comparison to highly accurate data~\cite{Dzhioev_2011} of experiments,
especially in interesting cases of low device-lead coupling.~\cite{PhysRevLett.95.146402} This overestimation can often
be attributed to too simplistic treatment of electron-electron
interactions.~\cite{PhysRevLett.93.036805,Strange2008} In the worst case, when many-body interaction effects
beyond mean-field become important these methods might even predict 
non-physical results such as low conductivity when the device would really show transparent conduction due to many-body effects.~\cite{PhysRevB.79.085120}

% HOW THEORY CAN CONTRIBUTE - MANY-BODY: WHAT IS THE PROBLEM?
Therefore it is desirable to extend these techniques with methods to tackle effects of strong correlations in an out of equilibrium context~\cite{Baer2003459, PhysRevB.77.115333,PhysRevB.80.115107,PhysRevB.79.155110,PhysRevB.80.165305,Dzhioev_2011,Arrigoni2012}. Many body effects of electron-electron interaction can be studied making use of simplified model Hamiltonians. By using a model based approach however, results may strongly depend on the model parameters which are usually notoriously difficult to estimate. For benzene, physically relevant interaction parameters have been obtained in literature by fitting experimental spectra~\cite{Bursill1998305, barford_2005} which provide the basis for many theoretical works.~\cite{PhysRevB.77.201406,PhysRevB.79.235404} Such parametrizations are typically not unique since considering different interaction matrix elements may yield again a good agreement with a specific set of experimental data but might promote very different physical mechanisms. Another approach, 
yielding model parameters is to determine them in an {\em ab-initio} way from first-principles calculations.~\cite{PhysRevLett.90.076805, PhysRevB.88.085404} Again these parameters are subject to changes due to screening effects, when the molecular device is finally coupled to leads or embedded in an environment.~\cite{doi:10.1021/nl8021708,kubatkin_single-electron_2003}

% OUR WORK: WHICH PROBLEM IS ADRESSED AND HOW
In this work we study many-body effects of electron-electron interaction beyond mean-field in the steady-state (sts) of a ring shaped molecular junction under voltage bias in a magnetic field. A simple Pariser-Parr-Pople model~\cite{Pariser_1953,Pople_1953} description for the atomic carbon $p_z$-orbitals is employed for a charge-neutral device coupled to two metallic leads. We address the question of how electron-electron interactions influence device properties in a magnetic field and identify their fingerprints. We do so by assessing naturally important quantities for molecular junctions, which are sts observables under applied bias voltage. We focus on studying the dependence of the sts currents, sts charge density as well as sts magnetization on electron-electron interactions and a magnetic field as well as their dependence on device-lead coupling. The purpose of this work is to concentrate on the qualitative effects of electronic correlations, rather than seeking quantitative agreement with experiments.
 A detailed, quantitative comparison to experiments would of course require the inclusion of additional electronic bands, mechanical vibrations, as well as charging and temperature effects. On the other hand correlation effects would be very difficult to treat at a comparable level of accuracy in such a rich model. In this work we focus on interaction effect on the most simple model for a benzene junction to identify basic mechanisms. The presented approach can however easily be generalized to include anchor groups or specific lead geometry.

% THEORETICAL APPROACH
To solve the interacting system out of equilibrium, we use cluster perturbation theory~\cite{balzer_non-equilibrium_2011,knap_nonequilibrium_2011,PhysRevB.86.245119}, which
amounts in treating the inter-cluster hopping within first order strong coupling perturbation theory. This approach is, in principle, refineable by considering larger cluster sizes.
Here, we target directly the sts which is formulated in the framework of sts Cluster Perturbation Theory (stsCPT). We work out general expressions for the sts density matrix and all bond currents in presence of a magnetic field in terms of stsCPT single-particle Green's functions. This framework provides a flexible, versatile and easy to use method for evaluating sts observables for interacting molecular junctions out of equilibrium in magnetic fields. It can  easily be generalized to other junctions and can straight forwardly be combined with {\em ab-initio} calculations.

% OUR RESULTS
Our results indicate that while properties of the electronic structure of the bare device including symmetry considerations and degeneracies have a dominant influence~\cite{PhysRevB.79.235404} on the sts behaviour, electron-electron interactions effectively renormalize conduction characteristics like threshold voltages and plateau currents in magnetic fields. We find that the sts charge density becomes strongly renormalized beyond mean-field and identify signals in the sts magnetization. This paves the way for even more sound and trust-able interpretation of results from sophisticated {\em ab-initio} based methods as well as model based calculations at fixed model parameters. Furthermore we re-examine effects introduced by magnetic fields in context with electron-electron interactions, lead induced broadening as well as lead electronic structure.

% OUTLINE
This paper is organized as follows. We present the theoretical model in \se~\ref{sec:benzene} and introduce the formalism for our calculations in \se~\ref{sec:method}. We outline how to systematically improve results obtained with the presented method and compare obtained data with existing calculations in \se~\ref{ssec:approx1}. Results for the sts currents, sts charge-density and sts magnetization are provided in \se~\ref{sec:results}.

\section{Model}\label{sec:benzene}
\begin{figure}
\includegraphics[width=0.45\textwidth]{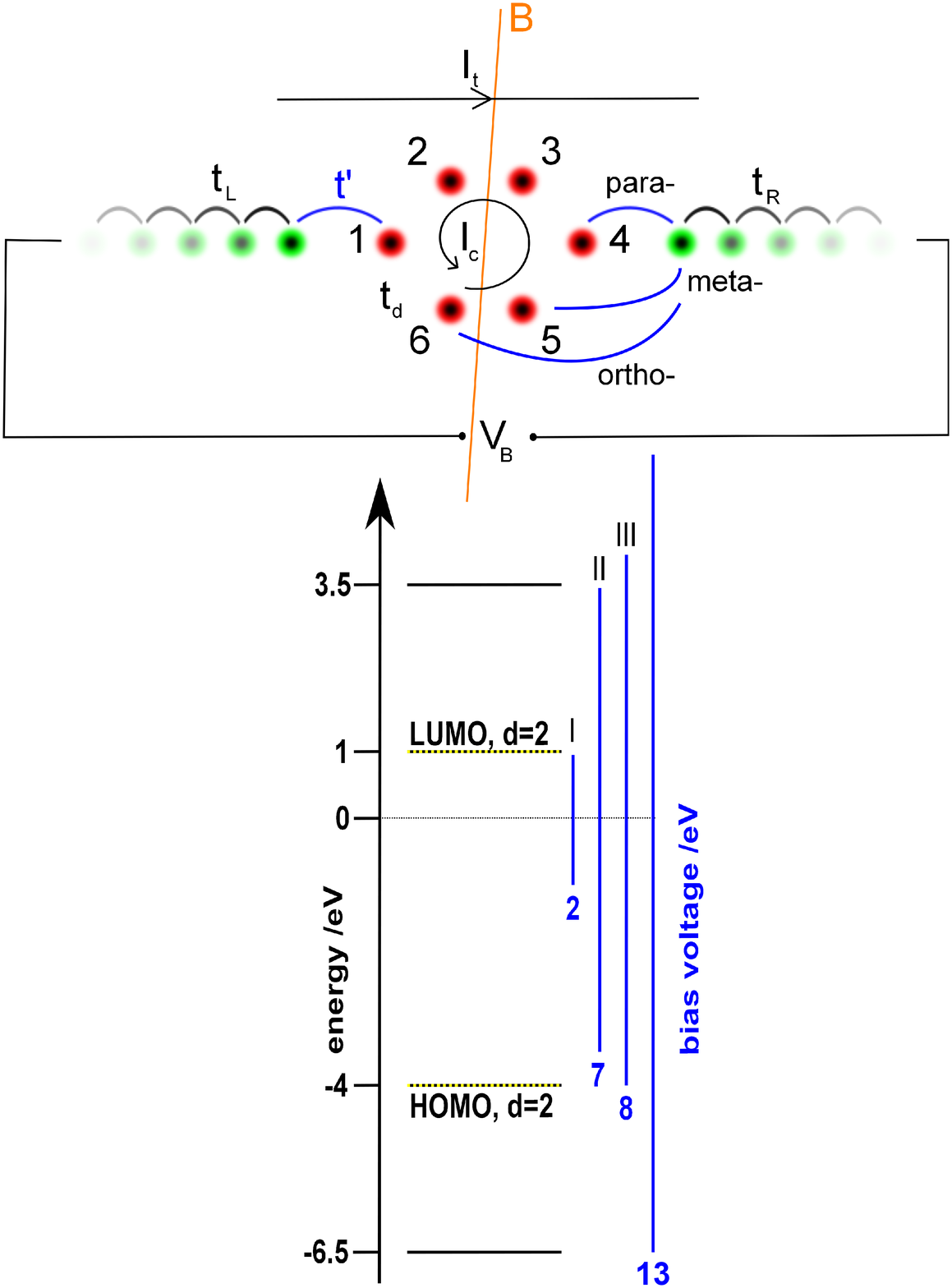}
\caption{(Color online) (top) Illustration of the device setup: A
  planar molecular ring is connected to two metallic reservoirs. The
  setup is placed in a perpendicular magnetic field $B$. A bias
  voltage $V_B$ is applied between the left and right lead. (bottom)
  Bare single-particle energy levels of the noninteracting,
  disconnected molecule. Bias voltages required for the specific
  levels to contribute to transport are indicated on the right. The
  HOMO as well as the LUMO are doubly degenerate ($d=2$) for $B=0$.}
\label{fig:ring}
\end{figure}
% OVERALL MODEL
We consider the effect of electron-electron interactions on the electric transport through a benzene like aromatic molecular ring. A simple starting point is provided by a tight-binding approach for the molecule coupled to left and right metallic leads (see \fig{fig:ring} (top)) 
\begin{align}
 \hat{\mathcal{H}} &= \hat{\mathcal{H}}_{\text{ring}} +\hat{\mathcal{H}}_{\text{lead}}^L +\hat{\mathcal{H}}_{\text{lead}}^R + \hat{\mathcal{H}}_{\text{ring-leads}}\,\mbox{.}
\label{eq:H}
\end{align}
% MOLECULAR MODEL
The benzene ring is modelled by considering one atomic $p_z$-orbital $\phi_{i\sigma}$ per carbon atom, yielding a six atomic-orbital Pariser-Parr-Pople~\cite{Pariser_1953,Pople_1953} (/(extended) Hubbard~\cite{Hubbard_1963}) Hamiltonian 
%\begin{subequations}
\begin{align}
\nonumber \hat{\mathcal{H}}_{\text{ring}}&=\sum\limits_{\sigma}\Bigg((\epsilon_d+\sigma\frac{B}{2}-\frac{U}{2}-2W)\sum\limits_{i=1}^{6}\,\hat{n}_{i\sigma}\\
\nonumber &+t_d\sum\limits_{i}^6\, \left(e^{i\Phi(B)}d_{i\sigma}^\dag d_{i+1\sigma}^\nag + \,e^{-i\Phi(B)}d_{i+1\sigma}^\dag d_{i\sigma}^\nag\right)\Bigg)\\
&+U\sum\limits_{i=1}^{6}\,\hat{n}_{i\uparrow}\hat{n}_{i\downarrow}+W\sum\limits_{\sigma\sigma'}\sum\limits_{<ij>}\,\hat{n}_{i\sigma}\hat{n}_{j\sigma'}\,\mbox{.}
\label{eq:Hring}
\end{align}
%\end{subequations}
where $i,j \in [1,6]$ enumerate the six ring orbitals in a clockwise fashion. The six nearest-neighbour bonds are denoted $<ij>$. Elementary fermionic operators $d_{i\sigma}/d_{i\sigma}^\dag$ annihilate/create an electron on the ring orbital $\phi_{i\sigma}$ with spin $\sigma=\{\uparrow, \downarrow\}$ and the particle number operator is defined in the usual way as $\hat{n}_{i\sigma}=d_{i\sigma}^\dag d_{i\sigma}$.

% MODEL PARAMETERS
The on-site energy of the ring orbitals is comprised of the bare on-site energy $\epsilon_d=-1.5\,$eV~\cite{doi:10.1021/jp105030d} (with respect to the leads), a Zeemann term $\sigma\frac{B}{2}$ and a correction cancelling the mean-field contribution of the on-site interaction $U$ and the nearest-neighbour density-density interaction $W$: $-\frac{U}{2}-2W$. Literature~\cite{Bursill1998305,barford_2005} provides an optimal parametrization for \eq{eq:Hring} when not connected to leads. Fitting to excitation spectra~\cite{NIST} the authors of \tcite{Bursill1998305} find that benzene is best described by a nearest-neighbour overlap integral of $t_d\approx-2.5\,$eV and an on-site {interaction} of $U\approx10\,$eV.~\cite{footnote2} Such a model is frequently used in literature~\cite{PhysRevB.77.201406,PhysRevB.79.235404} along with approaches where the interaction parameters are determined in an {\em ab-initio} way from first-principles calculations.~\cite{PhysRevLett.90.076805, PhysRevB.88.085404} We expect the 
values of 
$U$ and $W$ to be substantially reduced when the molecule is connected to the leads, due to screening from the band electrons.~\cite{doi:10.1021/nl8021708,kubatkin_single-electron_2003} Therefore we set in the following $t_d=-2.5\,$eV and discuss values of $U\leq9\,$eV as well as $W\leq3\,$eV. 

% MAGNETIC FIELD
We consider the effects of a magnetic field $B$ which is applied
perpendicular to the plane spanned by the molecular ring. 
This is described by a Peierls phase~\cite{footnote11} $\Phi(B)$ as well as by a Zeemann term. 

% MODEL LEADS
For simplicity, we model the right ($\alpha=R$) and left ($\alpha=L$) leads by semi-infinite tight-binding chains
\begin{align}
\hat{\mathcal{H}}_{\text{lead}}^\alpha&=\sum\limits_{\sigma}\Bigg((\epsilon_\alpha\pm\frac{V_B}{2})\sum\limits_{i=1}^{\infty}\,\hat{n}_{i\alpha\sigma}+t_\alpha\sum\limits_{<ij>}\, c_{i\alpha\sigma}^\dag c_{j\alpha\sigma}^\nag \Bigg)\,\mbox{,} \label{eq:Hlead}
\end{align}
where the fermionic operators in the lead orbitals $c_{i\alpha\sigma}^\nag/c_{i\alpha\sigma}^\dag$ are defined in a standard way. The resulting semi-circular electronic density of states (DOS) is centered around the leads' on-site energy $\epsilon_\alpha$ which is fixed to zero for zero bias. Application of a bias voltage $V_B$ is done by shifting the lead's chemical potential  (strictly speaking, at infinity) and on-site
energies to $\epsilon_\alpha=\mu_\alpha=\mp\frac{V_B}{2}$, so that the leads are kept half filled. We use a large lead nearest-neighbour hopping $t_\alpha=-6.0\,$eV (independent of $\alpha$), which implies a quite large bandwidth $D=24\,$eV, so that most of our results are comparable to the wide-band limit.~\cite{footnote10,doi:10.1021/jp105030d} We will also discuss and compare to differently shaped lead DOS when effects on the sts properties are to be expected.

The coupling between the ring and the leads is described by a single-particle hopping $t'$ to the left and right lead. The leads are connected to the molecule in the so-called para($1$,$4$), meta($1$,$5$) or ortho($1$,$6$) configuration. Here the two numbers in the braces label the ring position at which the (left, right) lead is attached, denoted ($1$, $x\in\{4, 5, 6\}$) in the following
\begin{align}
\nonumber \hat{\mathcal{H}}_{\text{ring-leads}}&=\sum\limits_{\sigma}\Bigg(t'\left(d_{1\sigma}^\dag c_{1L\sigma}^\nag + c_{1L\sigma}^\dag d_{1\sigma}^\nag\right)\\
&+t'\left(d_{x\sigma}^\dag c_{1R\sigma}^\nag + c_{1R\sigma}^\dag d_{x\sigma}^\nag\right)\Bigg)\,\mbox{.} \label{eq:Hring-lead}
\end{align}
This molecular-lead setup provides a good description of, e.g., 
two widely used experimental techniques: scanning tunnelling microscopy (STM)~\cite{venkataraman_dependence_2006} or MCBJ~\cite{Reed10101997}. Then the STM-tip/substrate material or break junction material enters in $\hat{\mathcal{H}}_{\text{lead}}$ through the DOS, while $\hat{\mathcal{H}}_{\text{ring-leads}}$ describes the tunnelling from the molecule into the experimental device. In most of the present work, we focus on a typically small value for the molecule-lead coupling of $t'=-0.05\,$eV~\cite{PhysRevB.85.155440} which enhances both the effects of electronic correlations and of the magnetic field. This leads to a level broadening of the order of $\Gamma \propto \pi t'^2 \text{DOS}_{\text{leads}}(\omega=0) \approx \frac{t'^2}{|t|} \approx 10^{-4}\,$eV. To study the effects of this lead induced broadening we also present results for 
larger $t'$.

% FINAL REMARKS
For the sake of simplicity we neglect effects of mechanical vibrations: The coupling of the electronic degrees of freedom to vibrations can lead to a renormalization of conductance thresholds as well as current plateaus as discussed in \tcite{knap_2012} and will be addressed in more detail in future work.~\cite{sorantin2013} Furthermore we do not discuss charging effects due to the connection of the leads.

\section{Method}\label{sec:method}
% WHAT IS DONE?
We aim at calculating sts properties of an interacting molecular device. Here we focus on the current between each bond as well as the charge distribution and magnetization.

% CURRENT CIRC
We consider two currents of special interest: i) the total transmission current $j_t$ and ii) the circular current $j_c$. In the presence of a bias-induced transmission current 
the two ring directions carry different currents. This leads, in principle, to an
ambiguity in the  definitions of the circular current. However, as
discussed by Rai \etal,~\cite{doi:10.1021/jp105030d} the most natural
expression is the one which is directly related to the current-induced
magnetic flux through the ring. According to Biot-Savart's
formula,~\cite{jackson} this is given by the average current obtained
by weighting the current flowing through each segment by its
length. In our case, in which there are two contacts dividing two ring
segments $i=1,2$ of lengths $L_i$ (in units of the lattice constant),
in which currents $\bar j_i$ flow (say, clockwise), we have 
\begin{align}
  j_c &=\frac{\left(\bar{j}_1 L_1+\bar{j}_2 L_2\right)}{6}\,\mbox{.}
\label{eq:jc}
\end{align}
In this work, we consider couplings to the leads in para: $L_1=3, L_2=3$, meta: $L_1=4, L_2=2$ and ortho: $L_1=5, L_2=1$ configuration (see \fig{fig:ring} (top)).

% CURRENT TRANS
The total transmission current is given by
\begin{align*}
  j_t &= \bar{j}_2-\bar{j}_1\,\mbox{,}
\end{align*}
which equals the in- as well as the outflow at the bonds connecting the leads to the ring by virtue of the continuity equation.

% DENSITY 
The sts charge distribution $\langle n \rangle$ and magnetization $\langle m \rangle$ can be obtained from the sts single-particle density-matrix $D_{ij}^\sigma$, which also encodes all information about the sts current. 

\subsection{Non equilibrium Green's functions}\label{ssec:negf}
% NEGF
One way of obtaining the sts density matrix is by making use of non equilibrium Green's functions in the Keldysh-Schwinger~\cite{schwinger_brownian_1961, feynman_theory_1963,keldysh_theory_1965} formalism 
\begin{align*}
  \widetilde{G} &= \begin{pmatrix} G^R & G^K \\ 0 & G^A \end{pmatrix}\;\mbox{,}
\end{align*}
where $G^{R/A/K}$ are matrices in orbital/spin space and functions of
two time coordinates $\tau$.~\cite{rammer_quantum_1986} $R$ denotes the
retarded, $A$ the advanced and $K$ the Keldysh component of the
single-particle Green's function in Keldysh space $\widetilde{G}$. Since we focus on the sts, time translation invariance applies and we can express the Green's functions in frequency space $\omega$.

% DENSITY MATRIX
The sts density matrix $D_{ij}^{\sigma}$ is obtained from the sts Green's function $\widetilde{G}(\omega)$ in matrix notation
\begin{align}
D_{ij}^{\sigma} &= \frac{\delta_{ij}}{2}-\frac{i}{2}\int\limits_{-\infty}^\infty\frac{d\omega}{2\pi}G^{K\sigma}_{ij}(\omega) \label{eq:D}\,\mbox{.}
\end{align}
% CURRENT
The sts current through orbital $i$ in presence of a magnetic field $B$ can be obtained by the time derivative of the total particle number of orbitals one either side of $i$.~\cite{haug_quantum_1996} In terms of equal time~\cite{footnote3} correlation functions one can express the current by Keldysh Green's functions $G^K=G^<+G^>$ in a symmetrized way $j_{i,i+1}^{\sigma} = \frac{e}{2\hbar}\Bigg(t_{i,i+1\sigma}G^{K\sigma}_{i,i+1}(\tau,\tau)-t_{i+1,i\sigma}G^{K\sigma}_{i+1,i}(\tau,\tau)\Bigg)$.

Generalizing the notation to arbitrary indices and keeping in mind that a definition of current only makes sense for nearest-neighbor orbitals, an expression in terms of the sts density matrix $D$ becomes available
\begin{align*}
j_{ij}^{\sigma} &=\frac{e}{2\hbar}\left( h_{ij}^{\sigma}\int\limits_{-\infty}^\infty\frac{d\omega}{2\pi}G^{K\sigma}_{ij}(\omega)-h_{ji}^{\sigma}\int\limits_{-\infty}^\infty\frac{d\omega}{2\pi}G^{K\sigma}_{ji}(\omega)\right)\\
&=\frac{i e}{\hbar}\left(h_{ij}^{\sigma}D_{ij}^{\sigma} - h_{ji}^{\sigma}D_{ji}^{\sigma}\right)\mbox{,}
\end{align*}
where $h_{ij}^{\sigma}$ denotes the single-particle part of $\hat{\mathcal{H}}$ (\eq{eq:H}) in the orbital/spin basis. The expression is purely real because $D$ and $h$ are hermitian.

Since we are concerned with electron-electron interactions, evaluating the needed Green's functions $\widetilde{G}$ is not possible in general.

\subsection{Steady-state Cluster Perturbation Theory}\label{ssec:ncpt}
% WHICH METHOD
We employ stsCPT~\cite{balzer_non-equilibrium_2011,gros_cluster_1993,senechal_spectral_2000} as outlined in \tcite{knap_nonequilibrium_2011, PhysRevB.86.245119} to construct an approximate solution for $\widetilde{G}(\omega)$ in the sts.

% GNERAL SCHEME
Within this approach the thermodynamically large system \eq{eq:H} is {split} into individually, exactly solvable parts at time $\tau\rightarrow-\infty$. The single-particle Green's function $\widetilde{g}(\omega)$ for each of these parts (clusters) is obtained by analytic or numeric means. The coupling between these parts is switched on at a later time $\tau_0$ using the inter-cluster (perturbation) matrix $T$, which holds the couplings between the disconnected parts. The sts Green's function of the full system in the stsCPT~\cite{gros_cluster_1993,senechal_spectral_2000} approximation is given by
\begin{align}
 \widetilde{G}(\omega)^{-1}&=\widetilde{g}(\omega)^{-1}-\widetilde{\uu}\otimes T\,\mbox{,}
\label{eq:CPT}
\end{align}
where $\widetilde{\uu}$ is the identity in Keldysh space. 
We use lower case $g$ for the single-particle Green's function of the initially decoupled equilibrium system while upper case $G$ denote the sts Green's functions of the fully coupled system. The stsCPT approximation made here is to replace the self-energy $\Sigma_G$ of the full system by the self-energy $\Sigma_g$
of the cluster. This amounts to a first order strong coupling
perturbation theory in inter-cluster terms $T$. The appealing
aspect of this approximation is that it becomes exact in each one of
three different limits: (i) for $T=0$, (ii) for $U=0$, or, in
principle, (iii) for an infinite cluster.

% PARTICULAR APPLICATION TO OUR PROBLEM
In our case, the system at $\tau\rightarrow-\infty$, thus, consists
of the molecular ring $\hat{\mathcal{H}}_{\text{ring}}$
(\eq{eq:Hring}), disconnected from the two leads
$\hat{\mathcal{H}}_{\text{lead}}^\alpha$ (\eq{eq:Hlead}), i.e. $t'=0$.
One can compute the retarded single-particle Green's function $g^R$ of the interacting
ring by a standard Lanczos approach,~\cite{lanc.51,ba.de.87} and the one of the non
interacting leads analytically.~\cite{economou_greens_2010, PhysRevB.85.235107} The
advanced component is available by the identity
$g^A=(g^{R})^\dag$. The Keldysh component $g^K$ of the initial systems,  which are separately in equilibrium, can by obtained by the
relation~\cite{negele_quantum_1998}  
\begin{align}
 g^K(\omega)&=(g^R(\omega)-g^A(\omega))(1-2p_{\text{FD}}(\omega,\mu, \beta))\,\mbox{,}
\label{eq:Keld}
\end{align}
where $p_{\text{FD}}(\omega,\mu,\beta)=\frac{1}{e^{\beta (\omega-\mu)}+1}$ is the Fermi-Dirac distribution function, $\beta$ denotes the inverse temperature and $\mu$ the chemical potential.

%% STATE AT tau->-inf
The state of the molecular device at $\tau\rightarrow-\infty$ is half
filled and unpolarized for all parameters under discussion in this
work. While, in principle, the sts results should not depend
  on the initial state of the finite size central part, 
 the approximate nature of the calculation, leads to some
dependences on the initial temperature and  chemical potential of the central region in
 the presence of interactions.
In general, we take the zero temperature ground state as a starting ($\tau\rightarrow-\infty$)
point, which leads to very good results in most cases. There are some
exceptions, in particular when degenerate states are involved. 
Therefore, blocking effects as discussed in 
~\cite{PhysRevB.77.201406,PhysRevB.79.235404} cannot be observed within
our approach. However, these are expected to occur for $B=0$ up to very small fields only,~\cite{PhysRevB.79.235404} since $B$ breaks the degeneracy responsible for these effects.

% MORE DETAILS
It is sufficient to calculate two $6\times6$ matrix Green's functions
for the central molecule (six orbitals times spin) and four scalar
Green's functions representing the contacting orbital of the two leads
(times spin).  Making use of \eq{eq:CPT} these three initial systems
are perturbatively connected by the coupling Hamiltonian
(\eq{eq:Hring-lead}) in the single-particle basis to obtain the sts
single-particle Green's function in Keldysh space
$\widetilde{G}(\omega)$. 

\subsection{Evaluation of steady-state observables}\label{ssec:evaluation}
Using the $\widetilde{G}(\omega)$ obtained via stsCPT, we rewrite the expression for the sts density matrix \eq{eq:D} in the corresponding language 
\begin{align}
\nonumber D_{ij}^{\sigma} &=\frac{i}{2}\int\limits_{-\infty}^\infty\frac{d\omega}{2\pi}\left(G^{R\sigma}_{in}(\omega)P^\sigma_{n j}(\omega)-P^\sigma_{in}(\omega) G^{R\sigma*}_{j n}(\omega)\right)\\
&+\frac{i}{2}\int\limits_{-\infty}^\infty\frac{d\omega}{2\pi}\left(G^{R\sigma}_{in}(\omega) \left(\left[P^\sigma(\omega),T^\sigma\right]_{-}\right)_{n m} G^{R\sigma*}_{j m}(\omega)\right)\,\mbox{,}
\label{eq:Dcpt}
\end{align}
where $[A, B]_{-}$ denotes the standard commutator and Einstein's summation convention is implicit. The occupation matrix $P^\sigma(\omega)$ corresponds to the second part of \eq{eq:Keld} and is diagonal: $P_{ij}^\sigma(\omega)=2\delta_{ij} p_{\text{FD}}(\omega,\mu_i,\beta_i)$, where $\mu_i$, $\beta_i$ are the chemical potential and inverse temperature of site $i$ at $\tau\rightarrow-\infty$. This expression is used to evaluate all sts observables as defined in \se~\ref{ssec:negf}. 

%% EVALUATING ALL OF THIS
The integrals in \eq{eq:Dcpt} are nonzero only between $\mu_{\text{min}}$ to $\mu_{\text{max}}$ which is due to the $P$ dependence and renders the numerical evaluation of \eq{eq:Dcpt} much more favourable than directly evaluating \eq{eq:D}. For a numerical evaluation a high precision adaptive integration scheme is necessary,~\cite{footnote4} especially when it comes to resolving small differences in bond currents as induced by magnetic fields (see \eq{eq:jc}).

\subsection{Quality of the stsCPT approximation and systematic improvements}\label{ssec:approx1}
% WHAT IS DONE
In the following we assess the quality of the stsCPT approximation and discuss systematic improvements for the treatment of correlations. Upon testing we found these improvements to be not important for this particular setup. They can however be be of considerable importance in systems exhibiting a more complicated quantum mechanical ground state.~\cite{PhysRevB.86.245119} 

% LARGER SIZE
As discussed above, one can, in principle, systematically improve results by enlarging the central region to include a certain number of orbitals of the leads in addition to the molecular ring. This means that at $\tau\rightarrow-\infty$ the system is disconnected at some bond $t$ of the leads chain. This sets much higher computational demands because the Hilbert space of the interacting part of the system grows exponentially in system size. In order to assess the accuracy of our results, we have investigated the convergence by adding one to three sites of each lead to the molecule and found negligible deviations in the sts results. This also allows us to conclude that data presented in this work are accurate and no serious error is being made due to the above mentioned approximations. Other expansions in the framework of Keldysh perturbation theory are discussed in \tcite{PhysRevB.79.245125} in this context.

% VARIATIONAL
Another way of improving the approximation for the self-energy  is to include  a self-consistent feedback within the non equilibrium Variational Cluster Approach.~\cite{knap_nonequilibrium_2011,PhysRevB.86.245119} We find this to be a major improvement for \textquotedblleft gapless\textquotedblright$
$systems. In this work we consider a system which has a large HOMO-LUMO gap and we found the improvements introduced by the self-consistency less important. Therefore, in this work we restrict to the non-variational stsCPT. Moreover, we restrict to zero temperatures, although finite temperatures are easily accessed.

\begin{figure*}
\includegraphics[width=0.9\textwidth]{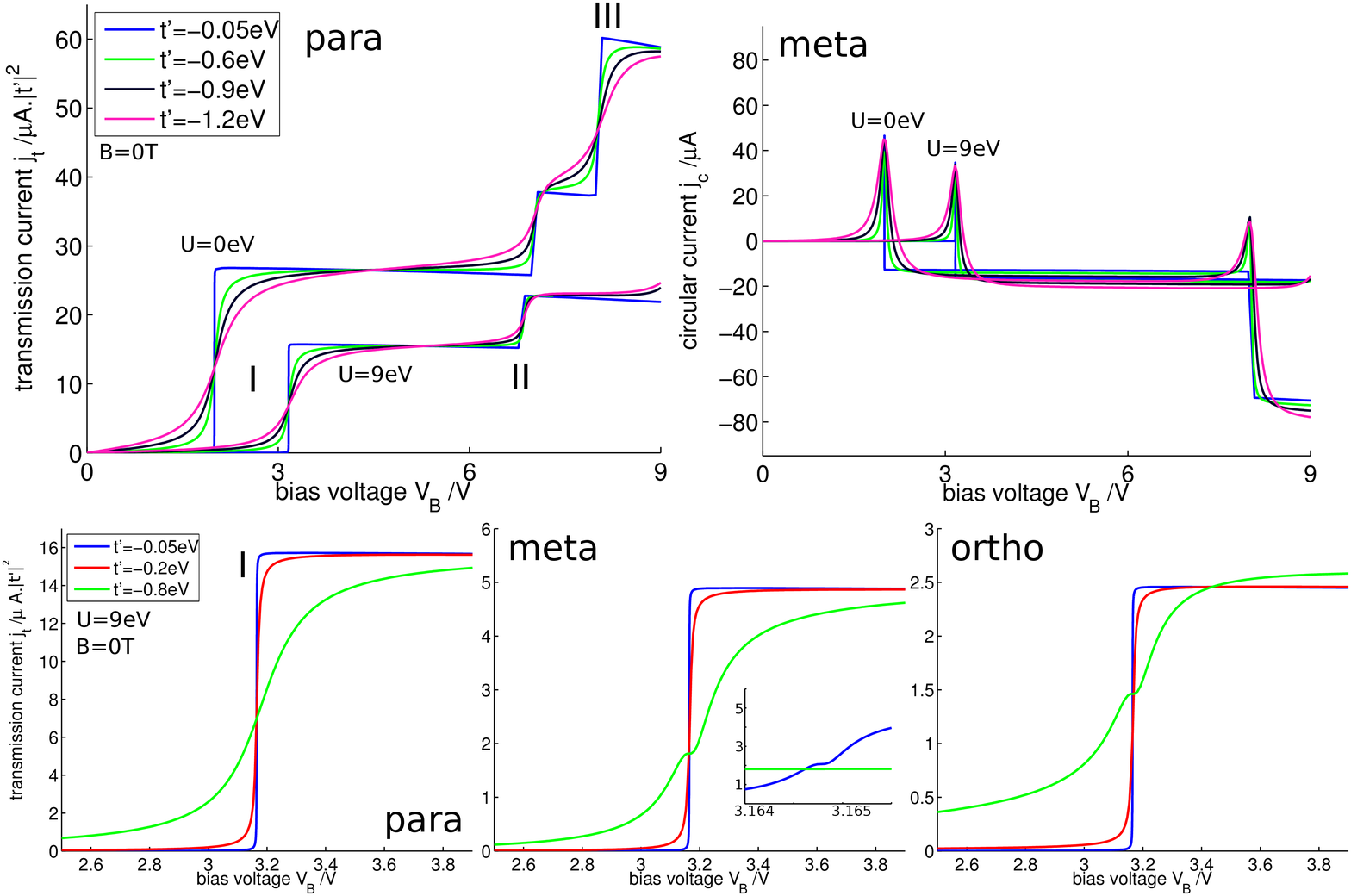}
\caption{(Color online) (top row) In the left panel, the total transmission current $j_t$ for molecule-lead couplings of $t'=-0.05\,$eV (blue), $t'=-0.6\,$eV (green), $t'=-0.9\,$eV (black) and $t'=-1.2\,$eV (magenta) is shown for para-connected benzene. In the right panel the circular current $j_c$ is shown for the same parameters. In each panel we visualize results for the noninteracting device and for an on-site interaction strength of $U=9\,$eV. (bottom row) The total transmission current $j_t$ for molecule-lead couplings of $t'=-0.05\,$eV (blue), $t'=-0.2\,$eV (red) and $t'=-0.8\,$eV (green) is shown for para-(left), meta-(center) and ortho-(right) connected benzene in the vicinity of the first threshold voltage. Data shown are for an on-site interaction strength of $U=9\,$eV in the zero field setup. The zoom in the central figure shows the ledge for $t'=0.05\,$eV in detail. Note that the vertical axis of the transmission current data is scaled by a factor of $\frac{1}{|t'|^2}$.}
\label{fig:ssCurrentTprime}
\end{figure*}

\section{Results and Discussion}\label{sec:results}
% OUTLINE
We present results for the sts properties of para-, meta- and ortho- connected benzene. We start by discussing lead induced broadening effects on the interacting ring within stsCPT for a broad range of molecule lead coupling strengths. Then we focus on the interesting case of small molecular lead coupling where effects of a magnetic field $B$ and electron-electron interactions become important. The current-voltage characteristics of the total transmission current $j_t$ and the circular current $j_c$ are presented, including a discussion of effects induced by the shape of the leads' DOS. We identify the behaviour of threshold voltages $V_T$ and maximum reachable currents. A detailed view on the highly interesting bias region of high variation in the current (around $V_T$) will be provided, identifying the effects of electron-electron interactions on the current signal shapes. Finally we examine the sts particle density $\langle n \rangle$ and magnetization $\langle m \rangle$ in the device.

% REFERENCE DATA BY OTHER GROUP FOR U=0
We note that the presented method becomes exact in the noninteracting limit $(U,W)\rightarrow 0$ and therefore has to agree with results for the noninteracting device presented by Rai \etal~\cite{PhysRevB.85.155440,doi:10.1021/jz200862r} obtained by calculations based on the Landauer formula~\cite{Landauer_1957} (not shown). We go beyond their detailed description in terms of transmission coefficients and QI effects by discussing effects of electron-electron interaction and by adding a Zeemann term to the discussion of the magnetic field. As discussed above, our approach may become unreliable in the presence of degeneracies in transport states, which occur in this model for $B=0$ (see \se~\ref{ssec:ncpt}). Therefore we do not observe the current blocking state which has been reported~\cite{PhysRevB.77.201406,PhysRevB.79.235404} in asymmetrically connected junctions at finite bias voltage for $U>0$ and $B=0$.

\subsection{Lead induced broadening}\label{ssec:Broadening}
% CPT + BROADENING
Lead induced broadening is obviously important for not too small values of the molecule-lead coupling $t'$, as we show in \fig{fig:ssCurrentTprime} (top). On the other hand, there are important broadening effects even for small $t'$, such as the ledge in the transmission current visible in \fig{fig:ssCurrentTprime} (bottom) for the meta and ortho setup, which is absent in the para case.~\cite{footnote8} It is clear that an accurate treatment of broadening effects is highly desirable, even more so when electron-electron interactions come into play. We find that the total transmission current scales with ${|t'|^2}$ while the circular current does not show a clear scaling as a function of $t'$.

% MASTER EQUATION
A technique commonly employed in the field of molecular electronics is based on generalized master equations.~\cite{PhysRevB.77.201406,PhysRevB.79.235404} Lead induced broadening is in general difficult to take into account accurately within such methods. stsCPT is able to capture broadening effects even in the interacting molecule, thus rendering stsCPT an interesting, complementary, approach to the generalized master equation technique.

%WHATS COMING
Note that we did not discuss the effects of a magnetic field here since they are fully washed away by the large $t'$ (for reasonable fields on the order of a few Tesla). Effects of electronic correlations become increasingly important for smaller $t'$. For these two reasons, in the rest of this work we focus on $t'=-0.05\,$eV because we aim at studying the interplay of a correlated device in a magnetic field.

\subsection{Field dependent current-voltage characteristics}\label{ssec:ssCurrentOverview}
\begin{figure*}
\includegraphics[width=0.9\textwidth]{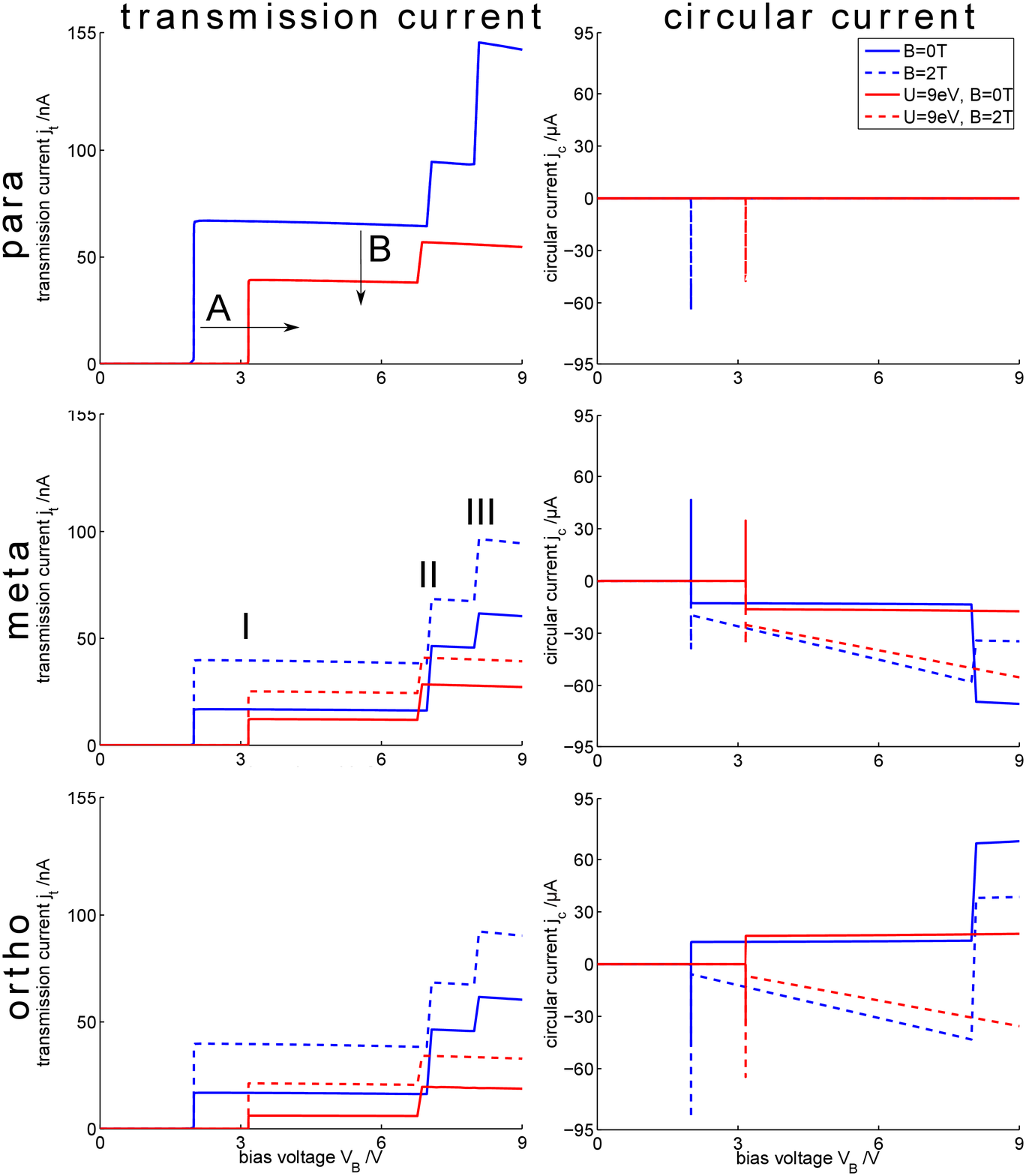}
\caption{(Color online) Current-voltage characteristics of para-(top), meta-(center) and ortho-(bottom) connected benzene. We plot the total transmission current $j_t /nA$ in the left and the ring current $j_c /\mu A$ in the right column. For every device setup, data for the noninteracting case (blue) is shown in comparison to results for an on-site interaction strength of $U=9\,$eV. We compare data for the zero field case (solid lines) to curves obtained for a magnetic field of $B=2\,$T (dash-dotted lines).}
\label{fig:ssCurrentOverview}
\end{figure*}
% WHAT IS THIS ABOUT
To ease navigation in the rest of the text we provide a bird's eye view on the current-voltage characteristics in a large voltage regime $V_B=[0, 9]\,V$ in \fig{fig:ssCurrentOverview}.  Here, effects of the finite bandwidth on the transmission current are still small. Since the magnetic field induces a very
small energy scale in comparison with the on-site, hopping and interaction energies in the Hamiltonian (\eq{eq:H}) we will be zooming into regions $V_B=\pm0.001\,V$ around selected voltage points. We note
that the absolute position on the bias-axis strongly depends on the
bare molecule-lead potential difference ($\epsilon_d=-1.5\,$eV) which
governs the first threshold voltage. 

% GENERAL STRUCTURE OF j AND THRESHOLD VOLTAGES
We start out by comparing the total transmission current $j_t$ and the
circular current $j_c$ of para-, meta- and ortho- connected benzene
for the noninteracting case as well as for $U=9\,$eV with and without
magnetic field $B$. $j_t$ shows multiple threshold voltages $V_T$ in a plateau like structure for all three device setups. At exactly these thresholds a signal spike in $j_c$ is observed. Those signals in $j_c$ are of considerable magnitude which are tens of $\mu$A, as compared to $j_t$ which is on the order of tens of nA, the later being mainly determined by the molecule-lead coupling $t'$. Such thresholds in $j_t$ / large signals in $j_c$ seem to be a generic feature of molecular ring devices~\cite{Nakanishi_1998,PhysRevLett.87.126801,doi:10.1021/ct600087c} and can be associated with nearly degenerate pairs of states with opposite angular momentum whose contribution to the net current is rendered small by destructive interference.~\cite{doi:10.1021/jp105030d} We discuss the interaction dependence of the threshold voltages $V_T$ in detail in \se~\ref{ssec:ssCurrentPosition}. Here we note that one effect of electron electron interaction is to shift the conduction thresholds $V_T$ (see A in \fig{fig:ssCurrentOverview}). Notice that this is a pure correlation effect, since the mean-field contribution has already been subtracted off.

% TRANSMISSION CURRENT
The magnitude of the plateaus in $j_t$ is fixed by the amount of current which can be carried by the molecular level starting to participate in transport at the corresponding $V_T$. Due to symmetry, $j_t$ in the para device is quite a bit higher than in the corresponding meta- or ortho- devices, the latter  two being comparable. In particular we find that the plateau current magnitudes are the same in the noninteracting meta- and ortho- devices but not when interactions are present (see also \se~\ref{ssec:ssCurrentMagnitude}). The transmission current of the para setup is independent of the magnetic field, while in the meta- and ortho- device $j_t$ grows until it seems to saturate at least for magnetic field magnitudes discussed in this work (see \se~\ref{ssec:ssCurrentMagnitude}). Increasing on-site interaction $U$ or density-density nearest-neighbor interaction $W$ (not shown) always decreases the total transmission current (see B in \fig{fig:ssCurrentOverview}, discussion in \se~\ref{ssec:ssCurrentMagnitude}). 

% RING CURRENT
The spikes in $j_c$ at $V_T$ (which are split by the Zeemann field, see \se~\ref{ssec:ssCurrentSignal}) are of considerable magnitude which is tens of $\mu A$. Similarly to $j_t$, increasing on-site interaction $U$ or nearest-neighbor density-density interaction $W$ reduces the magnitude of these signals. In the para device the ring currents exactly cancel for $B=0\,$T due to device symmetry while in the other two setups circular currents are also present in the field free system. Upon increasing the magnitude of the magnetic field all currents show saturation (see \se~\ref{ssec:ssCurrentMagnitude}). Note that for half filled systems in the large $U$ limit~\cite{footnote5} the current is unaffected by magnetic flux because electron motion becomes severely hindered.~\cite{Viefers20041} 

%INTERACTION EFFECTS
We find that effects of electron-electron interactions (shifting of conduction thresholds and decreasing plateau currents in $j_t$ and signal currents in $j_c$) are smooth and do not depend on the specific type of interaction (on-site, or nearest-neighbor density-density). 

\subsection{Effects originating from the electronic structure of the leads}\label{ssec:leadDOS}
% TRANSMISSION CURRENT
Instead of the often imposed wide-band DOS of the leads we use a semi-circular lead DOS with a large bandwidth ($t=-6\,$eV). As expected our setup mimics a wide-band limit regarding $j_t$ for low bias voltages where only a small bending down of the current can be observed due to the negative curvature at $\omega=0\,$eV of the leads' DOS (see \fig{fig:ssCurrentOverview} (left).

% RING CURRENT PARTICULAR
However, it turns out that $j_c$ is highly sensitive to even slight variations in the leads' DOS. Let
us first discuss the behaviour of $j_c$ in detail. In the para- setup,
the magnetic field induces peaks in $j_c$ at $V_T$, while in
between it falls back to plateaus in the nA range. In the meta- and
ortho- setup also at zero field a finite ring current exists which can
be amplified or suppressed by the magnetic flux, depending on its
direction according to Lenz's law.~\cite{jackson} These two setups
show a behaviour of $j_c$, which is different from the para
case. Without magnetic field, $j_c$ shows plateaus ($\propto V_B^2$)
in the $\mu$A range between the peaks. When a magnetic field is
imposed the two conducting states of the molecule at
$k=\pm\frac{2\pi}{3}$ and energy $\omega\approx 1\,$eV split up (see \fig{fig:ring} (bottom)). 
In our setup $j_c$ shows a linear growth between the signals starting
at the first threshold voltage. This large effect can be traced back
to different occupation of the $k=\pm\frac{2\pi}{3}$
states. Calculations for a constant lead DOS do not show this linear
increase in circular current. Instead constant plateaus (and therefore a
moderate imbalance in the population of the $k=\pm\frac{2\pi}{3}$
states) are obtained. These plateaus have a magnitude which $j_c$ acquires for the semi-circular DOS right after $V_T$. The crucial parameter here is the curvature of the
lead DOS which in the end amplifies the population imbalance and
renders it bias dependent. For a semi-circular DOS the coupling of the
two levels at $\omega\approx1\,$eV to the high bias and low bias lead
is of different magnitude (effectively $\Gamma$ becomes different for
the two leads) and this difference grows linearly with increasing bias
voltage. This effect has to our knowledge not been discussed before and is probably difficult to observe experimentally, especially since it strongly depends on the features of the DOS of the
leads. Moreover, electron-electron interaction does not play a role here. However, if a suitable system can be constructed, bias voltage could be used to linearly tune the circular current over a significant current magnitude.

\subsection{Current signal position}\label{ssec:ssCurrentPosition}
\begin{figure*}
\includegraphics[width=0.9\textwidth]{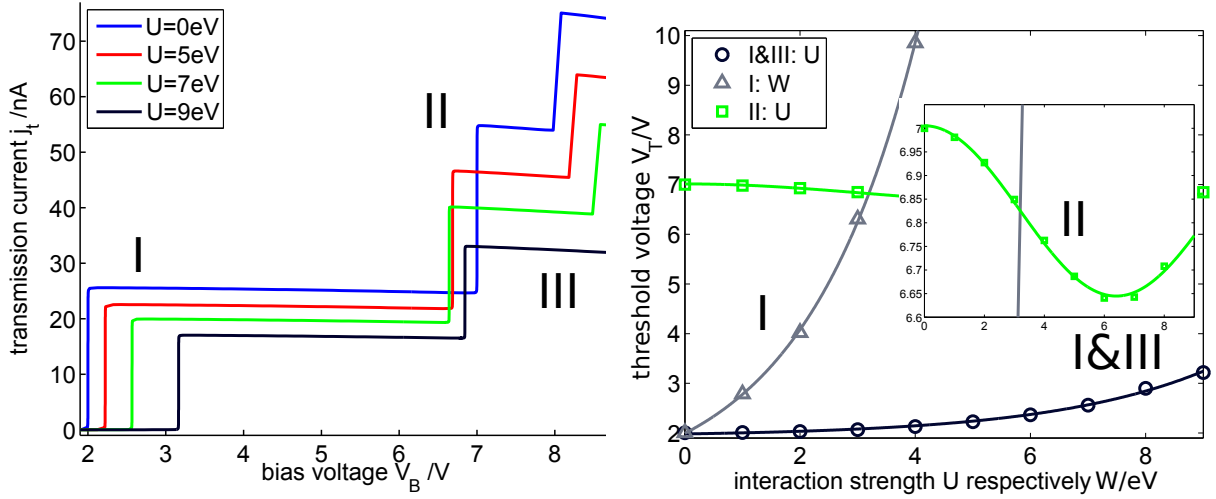}
\caption{(Color online) (left) Total transmission current $j_t$ as a function of bias voltage $V_B$ for various interaction strengths $U$. We illustrate data for meta-connected Benzene and a magnetic field of $B=1\,$T. (right) Threshold voltage $V_T$ at which the signal in the circular current $j_c$ /  inflection point in the transmission current $j_t$ occurs in region $I,II$ and $III$ (symbols).}
\label{fig:ssCurrentPosition}
\end{figure*}
% WHAT IS THIS
The transmission current-voltage characteristics consists of plateaus with steep jumps between them (see \fig{fig:ssCurrentPosition} (left)). These signal positions at threshold voltages $V_T$ are independent of the setup used (para-, meta- or ortho- connection) and furthermore independent of the magnetic field $B$.~\cite{footnote9} We plot data for meta-connected benzene for $B=1\,$T as a representative in \fig{fig:ssCurrentPosition} (left). The width of the current signals is $\approx 10^{-3}\,$eV (see \se~\ref{ssec:ssCurrentSignal}) due to the small lead-ring coupling $t'$.

% NON INTERACTING
For the noninteracting disconnected ring without magnetic field (\eq{eq:Hring}), the single particle energy levels are located at $\epsilon_k=\epsilon_d+2t_d \cos{(k)}=\{-6.5,-4,-4,1,1,3.5\}\,eV \text{ for } k=\frac{2\pi n}{6} \text{ and } n \in [-2,3]$, which leads to a ground state energy in the non polarized half filled case of $\omega_0=-29\,$eV with dynamics governed by a highest occupied molecular orbital (HOMO) - lowest unoccupied molecular orbital (LUMO) gap~\cite{footnote6} of $\Delta=5\,$eV (see \fig{fig:ring} (bottom)). Based on this energetic structure, one expects signals in the current at bias voltages of: $V_B\approx(2\,\times$ electronic level position$)$, i.e. at $\approx 2,7,8 \mbox{ and } 13\,$V. This is confirmed by our data (see \fig{fig:ssCurrentOverview}). 

% INTERACTING CASE
In the interacting device one can still interpret our data in terms of
the single-particle excitations of the molecular system obtained from
exact diagonalization of the molecular Hamiltonian
(\eq{eq:Hring}). Comparing the HOMO-LUMO gap $\Delta(U,W)$ to the data
for the sts current (see \fig{fig:ssCurrentPosition}) 
one finds a linear relation between the gap and the threshold voltage
$V_T^I=\Delta-3\,$eV. Note that the constant depends on the
position of  the molecular on-site energy with respect to the leads
 but the effects of interactions in
$\Delta$ are universal for weakly coupled charge neutral devices. From
our data we find a change in the HOMO-LUMO gap due to interactions
beyond mean-field $\Delta(U)[eV] = 5[eV] +
(0.02\pm0.01)\,e^{(0.43\pm0.03)U[eV]}$. The jump to plateau $III$
exhibits the same dependence on $U$ as the jump to plateau $I$ but
located at a different position:
$V_T^{III}=\Delta+3\,$eV. Interestingly enough these two thresholds are
monotonic in interaction strength but the threshold $II$ in between is
not. It shows a strongly non monotonic behaviour (see
\fig{fig:ssCurrentPosition}(right)).  

% FURTHER INTARXTING W
The functional dependence of the thresholds on on-site $U$ is the same as an nearest-neighbor density-density interactions $W$ (not shown). It is clear that the HOMO-LUMO gap grows much quicker with increasing $W$ than with increasing $U$.

% CONCLUDE INTERACTION EFFECTS
We conclude that the voltage thresholds in the parameter regime of the device under study which consists of i) small molecule-lead coupling $t'\ll t_d\ll t$ and ii) a charge neutral setting, are perturbatively accessible and can even be inferred to a high degree of accuracy from the single-particle spectrum of the isolated device. However, our results show that they can be subjected to non monotonic behaviour as a function of interaction strength. 

\subsection{Current signal magnitude}\label{ssec:ssCurrentMagnitude}
\begin{figure*}
\includegraphics[width=0.9\textwidth]{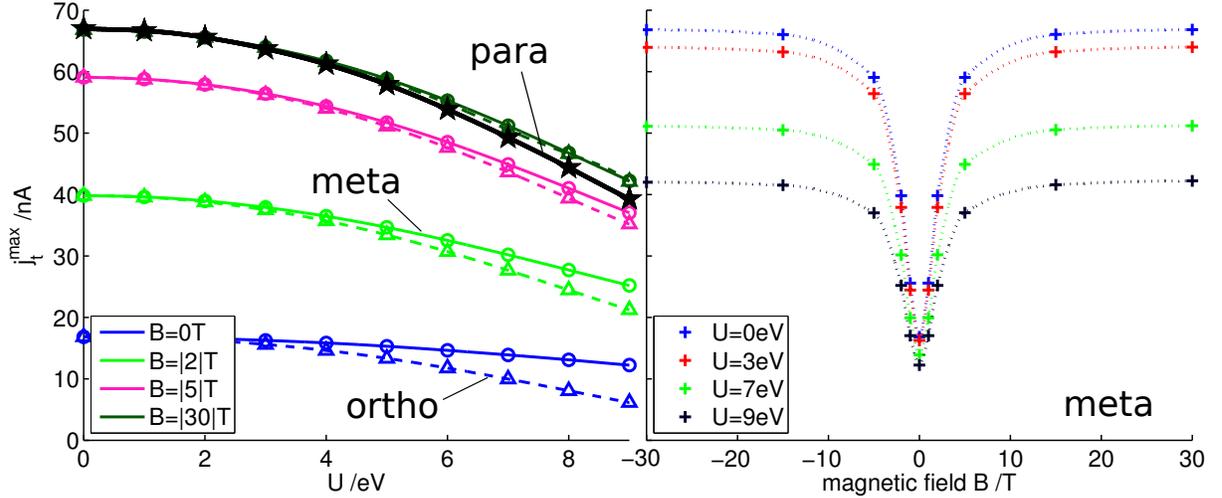}
\caption{(Color online) Maximal total transmission current $j_t^{\text{max}}$ in the vicinity of the first threshold voltage (I). (left) As a function of on-site interaction strength $U$ for the meta- (solid, circles) and ortho- (dashed, triangles) device. For the para- setup (thick black line, stars) it is independent of $B$ as a function of $U$. (right) The same quantity showing saturation as a function of $B$ for various values of on-site interaction strength $U$ in the meta-device.}
\label{fig:ssCurrentMaxima}
\end{figure*}
% WHAT IS THIS
We now turn to the analysis of the maximum current $j_t^{\text{max}}$ in the vicinity ($\pm0.1\,V$) of the first threshold voltage $V_T^I$.

% EFFECTS OF MAGNETIC FIELD
As noted before the para- setup does not show a magnetic field dependence in this channel for fixed interaction parameters due to device symmetry (thick black line, stars in \fig{fig:ssCurrentMaxima} (left)). In the meta- and ortho- devices, we find that the maximum current increases as a function of magnetic field $|B|$ until saturation for the considered range of magnetic field strength for all bias voltages (see \fig{fig:ssCurrentMaxima} (right)).

% DISCONNECTED DEVICE
Note that the eigenstates of the disconnected, noninteracting molecule in a magnetic field are $\epsilon_{n_\sigma} = \epsilon_d+\sigma\frac{B}{2}+2t_d \cos{(\frac{2\pi n_{\sigma}}{L}+\Phi(B))}$ with $n_\sigma\in[-2,3]$. The inherent circular current driven by the magnetic field is given by $j_c^{\text{inh}}=-\frac{e \hbar 2 \pi}{m_e 6^2}\sum\limits_{n_{\text{occ},\sigma}}n_\sigma$. From this expression it is clear that in this case the circular current magnitude is bounded from above because the magnetic field just redistributes occupied momenta of compact states. This analysis holds for all finite size quantum ring devices but not for the corresponding one dimensional field theory.~\cite{footnote7}

% EFFECTS OF INTERACTIONS
Depending on interaction strength $U$ we find that the maximum transmission current is monotonically decreasing (roughly $\propto \text{const.}-U^2$) in all device setups (see \fig{fig:ssCurrentMaxima} (left)). The same effect is observed for nearest-neighbor density-density interactions $W$ (not shown). The maximum transmission currents of the meta- and ortho- device are identical in the noninteracting device but start to differ when interactions are turned on.

\subsection{Current signal}\label{ssec:ssCurrentSignal}
\begin{figure*}
\includegraphics[width=0.9\textwidth]{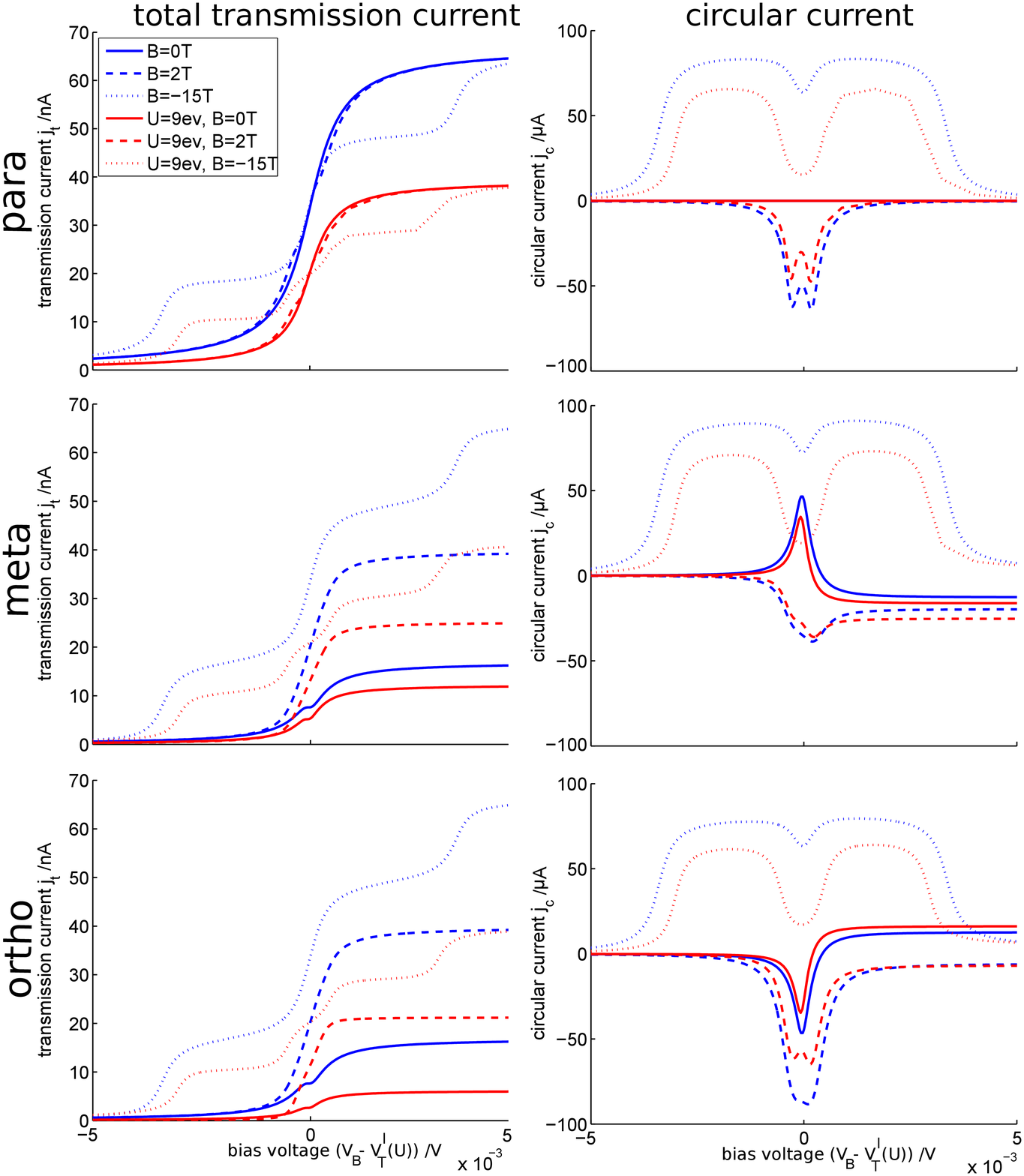}
\caption{(Color online) Total transmission current signal $j_t$ (left column) and ring current signal $j_s$ (right column) as a function of bias voltage $V_B$. From top to bottom we compare the current signal in the  para-, meta- and ortho- connected device for the noninteracting case $U=0\,$eV (blue) to an on-site interaction strength of $U=9\,$eV (red). We show data for zero field $B=0\,$T (solid), a field of $B=2\,$T (dashed) as well as $B=15\,$T(dotted). The signals have been shifted from their respective first threshold voltage (I) on top of each other to $V_B=0$.}
\label{fig:ssCurrentSignal}
\end{figure*}

% WHAT IS THIS
A zoom-in to the signal in the sts currents at the position of their respective first threshold voltages (I) is provided in \fig{fig:ssCurrentSignal}. We compare the currents for the noninteracting system with those of on-site interaction strength $U=9\,$eV. We observe similar signal shapes for all values of interaction strength ($U, W$). These signals are however shifted in bias depending on the electron-electron interaction parameters $(U,W)$ and the signal magnitude is decreased. In \fig{fig:ssCurrentSignal} we compare the signals of the noninteracting and interacting setups whereby the horizontal axis has been shifted in order for the threshold voltages to coincide.

% TRANSMISSION CURRENT
The total transmission current $j_t$ does not depend on the direction of the magnetic field in the para setup. The magnitude of the plateaus is however increased with increasing $|B|$ in the meta- and ortho- setup, while staying on the field free value in the para-setup. The effects of the Zeemann term are visible in $j_t$ due to a splitting $\propto |B|$ of the transition to the next plateau into two sub plateaus around the transition points $V_T$.

% CIRCULAR CURRENT
The circular currents $j_c$ in the para-setup depends on $B$ in a symmetric way i.e. reversing the direction of $B$ reverses just the sign of $j_c$. The meta-and ortho setups show a different magnitude in the circular current for the same $|B|$ due to inherent circular currents which are either amplified by the magnetic field or suppressed, depending on its alignment. The Zeemann splitting introduces a double peak structure around $V_T$ which is proportional to the magnetic field magnitude $|B|$. This splitting is introduced in a symmetric fashion around $V_T$. The circular currents $j_c$ grow with increasing magnetic field until a saturation is reached. Besides the decrease in circular current magnitude we observe a more pronounced Zeemann splitting with increasing interaction strength.

% INTERACTION
We again find a generic dependence on interactions beyond mean-field, i.e. using on-site or nearest-neighbor density-density interaction, which might even permit extracting interaction parameters for model Hamiltonians by carrying out transport measurements and fitting the position and height of sts currents at threshold voltages.
  
\subsection{Steady-state local charge density}\label{ssec:ssDensity}
\begin{figure*}
\includegraphics[width=0.9\textwidth]{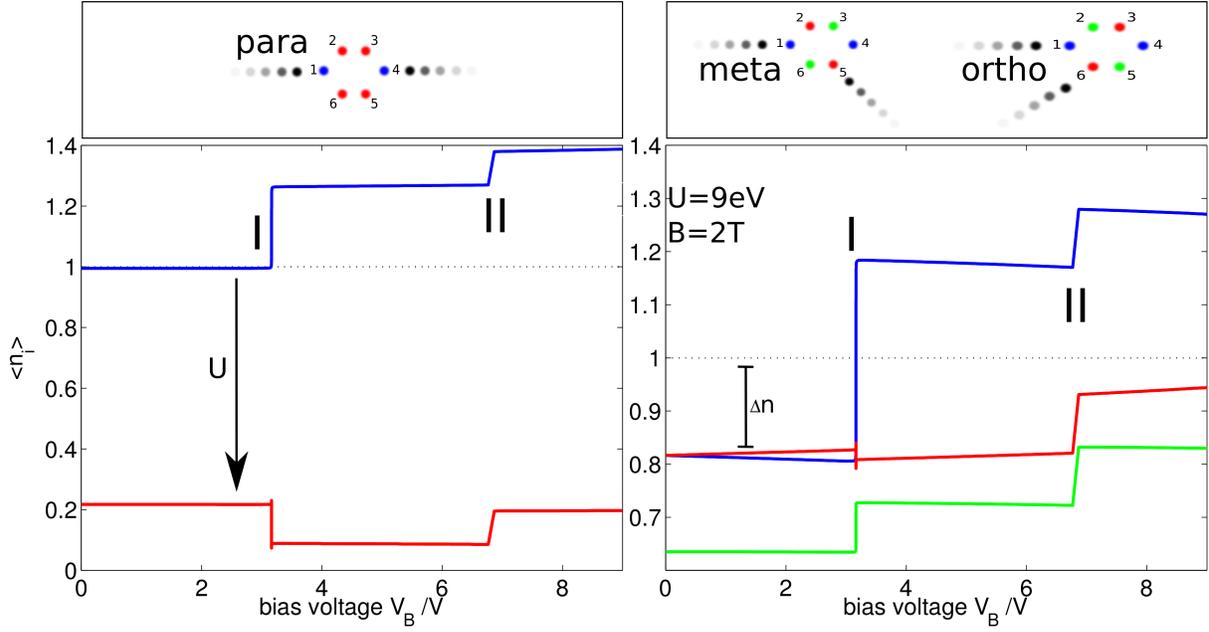}
\caption{(Color online) Ss local charge density of the para- connected device (left) as well as the meta- and ortho- connected devices (right). We show data for a device with on-site interaction of $U=9\,$eV in a magnetic field of $B=2\,$T. The drawings of the device at the top serve as a legend for the line colours to identify the respective orbitals. $\Delta n$ denotes the shift of the local density away from $\langle n_i \rangle=1$ in the low bias region.}
\label{fig:ssDensity}
\end{figure*}

% EFFECTS OF U
The sts charge density is shown in \fig{fig:ssDensity} for all three device setups for an interaction strength of $U=9\,$eV and a magnetic field of $B=2\,$T as an example for its generic behaviour. It exhibits features at the same threshold voltages as the sts current, where it either decreases or increases in a plateau like fashion. In the para setup (\fig{fig:ssDensity} (left)) we observe an increased charge density at the orbitals at which the leads are connected and a strongly reduced charge density in the other orbitals of the molecule due to interactions i.e. for the noninteracting molecule all densities would start at one. The sts charge density of the {meta- and ortho-} setups (\fig{fig:ssDensity} (right)) are symmetry related. They show the same sts charge density except for the fact that two orbitals exchange their respective charge density each.

% SUPPRESSION
The suppression of occupation is most pronounced in the orbitals in the ``interior'' of the molecule, respecting symmetry. These are orbitals $2,3,5,6$ in the para-device, orbitals $3,6$ in the meta device and orbitals $2,5$ in the ortho device.

% OTHER U
For smaller interaction-strength the overall shape of the curves does not change. However we find that in all setups, for on-site interaction strengths $U\leq5\,$eV, all local densities are one in the low bias regime up to the first threshold voltage: $\Delta n=0$, and depend on the interaction strength beyond that point. For interaction strength $U\geq6\,$eV, $\Delta n$ depends on interaction also in the low bias region.

% LARGER COUPLINGS
We note that for increasing molecule-lead coupling $t'$ the magnitude of the sts local charge density stays the same but the transitions become washed out (not shown). Furthermore with increasing bias voltage as well as in the vicinity of the threshold voltages a splitting of the degenerate orbitals occurs.

% CONCLUSION
The main effect of interactions beyond mean-field is to suppress the sts charge density respecting the symmetry of the isolated molecule. Such substantial charge redistribution in sts transport due to interactions is an important effect to be considered when discussing results from self-consistent DFT+NEGF calculations.

\subsection{Steady-state magnetization}\label{ssec:ssMagnetization}
\begin{figure*}
\includegraphics[width=0.9\textwidth]{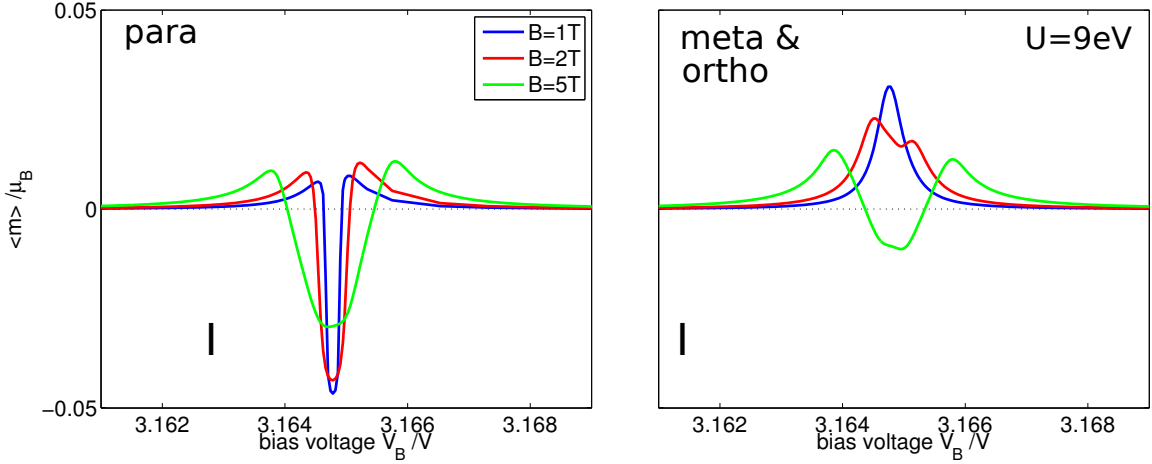}
\caption{(Color online) Signals in the sts magnetization of the device due to the Zeemann effect. We compare the effects of magnetic fields of $B=1\,$T (blue), $B=2\,$T (red) and $B=5\,$T (green) for a device with an on-site interaction of $9\,$eV.}
\label{fig:ssMagnetization}
\end{figure*}
% WHAT IS THIS
In our calculations a Zeemann term is included. Therefore the magnetic field induces a sts magnetization in the vicinity of the threshold voltages $V_T$. We consider just the paramagnetic (spin) contribution, since it is the dominant term. In \fig{fig:ssMagnetization}, we show data for all three device setups in the vicinity of the first voltage threshold (I) for the interacting case.
Increasing the magnetic field turns the resonance like structure in $j_c$ around $V_T$ into a structure which
consists of two maxima on the positive and one on the negative axis in between. The effect of changing the direction of the magnetic field while keeping $|B|$ constant is to mirror the sts magnetization curves around the x-axis (not shown). Without interactions the effect is quite similar and only of slightly different magnitude.

% CONCLUSION
We find that the sts magnetization is a highly sensitive quantity as a function of bias voltage (in the sub milli-volt regime) in the region around $V_T$ where it is possible to alter the direction as well as the magnitude of the magnetization by $\approx \pm10^{-2}/\mu_B$.  

\section{Conclusions}\label{sec:conclusions}
% WHAT WE DID
We study effects of electron-electron interactions in a benzene like
ring shaped molecule, subject to a finite bias voltage induced by two metallic leads,
as a function of an applied magnetic field.
We make use of a Hubbard-type model based description of such a device
in the charge neutral regime. Using steady-state Cluster Perturbation Theory, observables have been computed for para-, meta- as well as ortho- setups. 

% WHAT HAS BEEN PRESENTED
Results for the total transmission current and circular current as well as the steady-state charge
distribution and magnetization have been
presented. By studying physically relevant regimes of
electron-electron interactions in addition to an applied magnetic
field we describe the effects of electron-electron interactions on the steady-state beyond the mean-field level. We found that these are to shift voltage thresholds and to decrease
the magnitude of currents. Additionally, interactions lead to
deviating currents in the meta- and ortho- setup which were comparable
in the noninteracting system. The steady-state charge distribution becomes
strongly renormalized by interactions respecting the symmetry of the
isolated molecule. Due to the Zeemann effect we obtain a steady-state
magnetization which is highly sensitive to bias voltage. 

% IMPLICATIONS
Our results may help to validate model calculations at fixed interaction parameters and contribute to the understanding of sophisticated {\em ab-initio} based transport calculations. They might even contribute to designing empirical formulas for junction engineering. Our results indicate that the main effect of interactions is to renormalize voltage thresholds and current magnitudes. Care has to be taken in 
discussing symmetry relations of meta- and ortho- connected devices. Furthermore we showed that the charge density is sensitive to electron-electron interactions and becomes strongly renormalized with every additional electronic level contributing to transport. This fact has to be accounted for in self-consistent approaches. We find that insights into the relevant interaction parameters as well as their magnitude could be extracted from transport measurements, provided the on-site potential of the molecule with respect to the leads is known.

% THEORY
We presented general expressions for steady-state observables in
the language of steady-state Cluster Perturbation Theory. The
presented formalism is flexible and simple to apply to a broad range
of molecular junctions. Its approximate nature is systematically
improvable and the method does work in all parameter regimes. The
method is especially interesting in parameter regimes which exhibit
extremely high time scales rendering approaches based on time
evolution not feasible and when lead induced broadening effects become
important which render generalized master equation techniques
difficult. The presented approach is, however, unable
   to capture interaction mediated interference effects, like
  current blocking in asymmetrically connected junctions, when
  degeneracies in transport channels are present. This can be overcome
  by a more elaborate construction of the ``starting'' ($\tau\rightarrow
  -\infty$) cluster state. Work along these lines is in progress.

\begin{acknowledgments}
We gratefully acknowledge fruitful discussions with Karin Zojer, Dmitry A. Ryndyk and Max Sorantin. This work was partly supported by the Austrian Science Fund (FWF) P24081-N16 as well as SFB-ViCoM sub projects F04103 and F04104. MN thanks the MPI-PKS Dresden, in particular Masud Haque and the QSOE13 for hospitality.
\end{acknowledgments}

\bibliography{Benzene}{}

\begin{thebibliography}{115}
\expandafter\ifx\csname natexlab\endcsname\relax\def\natexlab#1{#1}\fi
\expandafter\ifx\csname bibnamefont\endcsname\relax
  \def\bibnamefont#1{#1}\fi
\expandafter\ifx\csname bibfnamefont\endcsname\relax
  \def\bibfnamefont#1{#1}\fi
\expandafter\ifx\csname citenamefont\endcsname\relax
  \def\citenamefont#1{#1}\fi
\expandafter\ifx\csname url\endcsname\relax
  \def\url#1{\texttt{#1}}\fi
\expandafter\ifx\csname urlprefix\endcsname\relax\def\urlprefix{URL }\fi
\providecommand{\bibinfo}[2]{#2}
\providecommand{\eprint}[2][]{\url{#2}}

\bibitem[{\citenamefont{Cuniberti et~al.}(2005)\citenamefont{Cuniberti, Fagas,
  and Richter}}]{cuniberti_2005}
\bibinfo{author}{\bibfnamefont{G.}~\bibnamefont{Cuniberti}},
  \bibinfo{author}{\bibfnamefont{G.}~\bibnamefont{Fagas}}, \bibnamefont{and}
  \bibinfo{author}{\bibfnamefont{K.}~\bibnamefont{Richter}},
  \emph{\bibinfo{title}{Introducing Molecular Electronics}}
  (\bibinfo{publisher}{Springer}, \bibinfo{year}{2005}), ISBN
  \bibinfo{isbn}{3540279946}.

\bibitem[{\citenamefont{Smit et~al.}(2002)\citenamefont{Smit, Noat, Untiedt,
  Lang, van Hemert, and van Ruitenbeek}}]{smit_measurement_2002}
\bibinfo{author}{\bibfnamefont{R.~H.~M.} \bibnamefont{Smit}},
  \bibinfo{author}{\bibfnamefont{Y.}~\bibnamefont{Noat}},
  \bibinfo{author}{\bibfnamefont{C.}~\bibnamefont{Untiedt}},
  \bibinfo{author}{\bibfnamefont{N.~D.} \bibnamefont{Lang}},
  \bibinfo{author}{\bibfnamefont{M.~C.} \bibnamefont{van Hemert}},
  \bibnamefont{and} \bibinfo{author}{\bibfnamefont{J.~M.} \bibnamefont{van
  Ruitenbeek}}, \bibinfo{journal}{Nature} \textbf{\bibinfo{volume}{419}},
  \bibinfo{pages}{906} (\bibinfo{year}{2002}).

\bibitem[{\citenamefont{Park et~al.}(2002)\citenamefont{Park, Pasupathy,
  Goldsmith, Chang, Yaish, Petta, Rinkoski, Sethna, Abruna, {McEuen}
  et~al.}}]{park_coulomb_2002}
\bibinfo{author}{\bibfnamefont{J.}~\bibnamefont{Park}},
  \bibinfo{author}{\bibfnamefont{A.~N.} \bibnamefont{Pasupathy}},
  \bibinfo{author}{\bibfnamefont{J.~I.} \bibnamefont{Goldsmith}},
  \bibinfo{author}{\bibfnamefont{C.}~\bibnamefont{Chang}},
  \bibinfo{author}{\bibfnamefont{Y.}~\bibnamefont{Yaish}},
  \bibinfo{author}{\bibfnamefont{J.~R.} \bibnamefont{Petta}},
  \bibinfo{author}{\bibfnamefont{M.}~\bibnamefont{Rinkoski}},
  \bibinfo{author}{\bibfnamefont{J.~P.} \bibnamefont{Sethna}},
  \bibinfo{author}{\bibfnamefont{H.~D.} \bibnamefont{Abruna}},
  \bibinfo{author}{\bibfnamefont{P.~L.} \bibnamefont{{McEuen}}},
  \bibnamefont{et~al.}, \bibinfo{journal}{Nature}
  \textbf{\bibinfo{volume}{417}}, \bibinfo{pages}{722} (\bibinfo{year}{2002}).

\bibitem[{\citenamefont{Liang et~al.}(2002)\citenamefont{Liang, Shores,
  Bockrath, Long, and Park}}]{liang_kondo_2002}
\bibinfo{author}{\bibfnamefont{W.}~\bibnamefont{Liang}},
  \bibinfo{author}{\bibfnamefont{M.~P.} \bibnamefont{Shores}},
  \bibinfo{author}{\bibfnamefont{M.}~\bibnamefont{Bockrath}},
  \bibinfo{author}{\bibfnamefont{J.~R.} \bibnamefont{Long}}, \bibnamefont{and}
  \bibinfo{author}{\bibfnamefont{H.}~\bibnamefont{Park}},
  \bibinfo{journal}{Nature} \textbf{\bibinfo{volume}{417}},
  \bibinfo{pages}{725} (\bibinfo{year}{2002}).

\bibitem[{\citenamefont{Agrait et~al.}(2003)\citenamefont{Agrait, Yeyati, and
  van Ruitenbeek}}]{Agrait200381}
\bibinfo{author}{\bibfnamefont{N.}~\bibnamefont{Agrait}},
  \bibinfo{author}{\bibfnamefont{A.~L.} \bibnamefont{Yeyati}},
  \bibnamefont{and} \bibinfo{author}{\bibfnamefont{J.~M.} \bibnamefont{van
  Ruitenbeek}}, \bibinfo{journal}{Physics Reports}
  \textbf{\bibinfo{volume}{377}}, \bibinfo{pages}{81 } (\bibinfo{year}{2003}).

\bibitem[{\citenamefont{Venkataraman
  et~al.}(2006{\natexlab{a}})\citenamefont{Venkataraman, Klare, Nuckolls,
  Hybertsen, and Steigerwald}}]{venkataraman_dependence_2006}
\bibinfo{author}{\bibfnamefont{L.}~\bibnamefont{Venkataraman}},
  \bibinfo{author}{\bibfnamefont{J.~E.} \bibnamefont{Klare}},
  \bibinfo{author}{\bibfnamefont{C.}~\bibnamefont{Nuckolls}},
  \bibinfo{author}{\bibfnamefont{M.~S.} \bibnamefont{Hybertsen}},
  \bibnamefont{and} \bibinfo{author}{\bibfnamefont{M.~L.}
  \bibnamefont{Steigerwald}}, \bibinfo{journal}{Nature}
  \textbf{\bibinfo{volume}{442}}, \bibinfo{pages}{904}
  (\bibinfo{year}{2006}{\natexlab{a}}).

\bibitem[{\citenamefont{Nitzan and Ratner}(2003)}]{Nitzan30052003}
\bibinfo{author}{\bibfnamefont{A.}~\bibnamefont{Nitzan}} \bibnamefont{and}
  \bibinfo{author}{\bibfnamefont{M.~A.} \bibnamefont{Ratner}},
  \bibinfo{journal}{Science} \textbf{\bibinfo{volume}{300}},
  \bibinfo{pages}{1384} (\bibinfo{year}{2003}).

\bibitem[{\citenamefont{Reed et~al.}(1997)\citenamefont{Reed, Zhou, Muller,
  Burgin, and Tour}}]{Reed10101997}
\bibinfo{author}{\bibfnamefont{M.~A.} \bibnamefont{Reed}},
  \bibinfo{author}{\bibfnamefont{C.}~\bibnamefont{Zhou}},
  \bibinfo{author}{\bibfnamefont{C.~J.} \bibnamefont{Muller}},
  \bibinfo{author}{\bibfnamefont{T.~P.} \bibnamefont{Burgin}},
  \bibnamefont{and} \bibinfo{author}{\bibfnamefont{J.~M.} \bibnamefont{Tour}},
  \bibinfo{journal}{Science} \textbf{\bibinfo{volume}{278}},
  \bibinfo{pages}{252} (\bibinfo{year}{1997}).

\bibitem[{\citenamefont{L\"ortscher et~al.}(2007)\citenamefont{L\"ortscher,
  Weber, and Riel}}]{PhysRevLett.98.176807}
\bibinfo{author}{\bibfnamefont{E.}~\bibnamefont{L\"ortscher}},
  \bibinfo{author}{\bibfnamefont{H.~B.} \bibnamefont{Weber}}, \bibnamefont{and}
  \bibinfo{author}{\bibfnamefont{H.}~\bibnamefont{Riel}},
  \bibinfo{journal}{Phys. Rev. Lett.} \textbf{\bibinfo{volume}{98}},
  \bibinfo{pages}{176807} (\bibinfo{year}{2007}).

\bibitem[{\citenamefont{Cuevas and Scheer}(2010)}]{Cuevas2010}
\bibinfo{author}{\bibfnamefont{J.~C.} \bibnamefont{Cuevas}} \bibnamefont{and}
  \bibinfo{author}{\bibfnamefont{E.}~\bibnamefont{Scheer}},
  \emph{\bibinfo{title}{Molecular Electronics: An Introduction to Theory and
  Experiment}} (\bibinfo{publisher}{World Scientific Publishing Company},
  \bibinfo{year}{2010}), \bibinfo{edition}{1st} ed., ISBN
  \bibinfo{isbn}{9814282588}.

\bibitem[{\citenamefont{Kiguchi et~al.}(2008)\citenamefont{Kiguchi, Tal,
  Wohlthat, Pauly, Krieger, Djukic, Cuevas, and van
  Ruitenbeek}}]{PhysRevLett.101.046801}
\bibinfo{author}{\bibfnamefont{M.}~\bibnamefont{Kiguchi}},
  \bibinfo{author}{\bibfnamefont{O.}~\bibnamefont{Tal}},
  \bibinfo{author}{\bibfnamefont{S.}~\bibnamefont{Wohlthat}},
  \bibinfo{author}{\bibfnamefont{F.}~\bibnamefont{Pauly}},
  \bibinfo{author}{\bibfnamefont{M.}~\bibnamefont{Krieger}},
  \bibinfo{author}{\bibfnamefont{D.}~\bibnamefont{Djukic}},
  \bibinfo{author}{\bibfnamefont{J.~C.} \bibnamefont{Cuevas}},
  \bibnamefont{and} \bibinfo{author}{\bibfnamefont{J.~M.} \bibnamefont{van
  Ruitenbeek}}, \bibinfo{journal}{Phys. Rev. Lett.}
  \textbf{\bibinfo{volume}{101}}, \bibinfo{pages}{046801}
  (\bibinfo{year}{2008}).

\bibitem[{\citenamefont{Song et~al.}(2009)\citenamefont{Song, Kim, Jang, Jeong,
  Reed, and Lee}}]{song_observation_2009}
\bibinfo{author}{\bibfnamefont{H.}~\bibnamefont{Song}},
  \bibinfo{author}{\bibfnamefont{Y.}~\bibnamefont{Kim}},
  \bibinfo{author}{\bibfnamefont{Y.~H.} \bibnamefont{Jang}},
  \bibinfo{author}{\bibfnamefont{H.}~\bibnamefont{Jeong}},
  \bibinfo{author}{\bibfnamefont{M.~A.} \bibnamefont{Reed}}, \bibnamefont{and}
  \bibinfo{author}{\bibfnamefont{T.}~\bibnamefont{Lee}},
  \bibinfo{journal}{Nature} \textbf{\bibinfo{volume}{462}},
  \bibinfo{pages}{1039} (\bibinfo{year}{2009}).

\bibitem[{\citenamefont{Kocherzhenko et~al.}(2011)\citenamefont{Kocherzhenko,
  Siebbeles, and Grozema}}]{doi:10.1021/jz200535j}
\bibinfo{author}{\bibfnamefont{A.~A.} \bibnamefont{Kocherzhenko}},
  \bibinfo{author}{\bibfnamefont{L.~D.~A.} \bibnamefont{Siebbeles}},
  \bibnamefont{and} \bibinfo{author}{\bibfnamefont{F.~C.}
  \bibnamefont{Grozema}}, \bibinfo{journal}{The Journal of Physical Chemistry
  Letters} \textbf{\bibinfo{volume}{2}}, \bibinfo{pages}{1753}
  (\bibinfo{year}{2011}).

\bibitem[{\citenamefont{Ke et~al.}(2008)\citenamefont{Ke, Yang, and
  Baranger}}]{doi:10.1021/nl8016175}
\bibinfo{author}{\bibfnamefont{S.-H.} \bibnamefont{Ke}},
  \bibinfo{author}{\bibfnamefont{W.}~\bibnamefont{Yang}}, \bibnamefont{and}
  \bibinfo{author}{\bibfnamefont{H.~U.} \bibnamefont{Baranger}},
  \bibinfo{journal}{Nano Letters} \textbf{\bibinfo{volume}{8}},
  \bibinfo{pages}{3257} (\bibinfo{year}{2008}).

\bibitem[{\citenamefont{Stadler et~al.}(2003)\citenamefont{Stadler, Forshaw,
  and Joachim}}]{0957-4484-14-2-307}
\bibinfo{author}{\bibfnamefont{R.}~\bibnamefont{Stadler}},
  \bibinfo{author}{\bibfnamefont{M.}~\bibnamefont{Forshaw}}, \bibnamefont{and}
  \bibinfo{author}{\bibfnamefont{C.}~\bibnamefont{Joachim}},
  \bibinfo{journal}{Nanotechnology} \textbf{\bibinfo{volume}{14}},
  \bibinfo{pages}{138} (\bibinfo{year}{2003}).

\bibitem[{\citenamefont{Rai et~al.}(2012)\citenamefont{Rai, Hod, and
  Nitzan}}]{PhysRevB.85.155440}
\bibinfo{author}{\bibfnamefont{D.}~\bibnamefont{Rai}},
  \bibinfo{author}{\bibfnamefont{O.}~\bibnamefont{Hod}}, \bibnamefont{and}
  \bibinfo{author}{\bibfnamefont{A.}~\bibnamefont{Nitzan}},
  \bibinfo{journal}{Phys. Rev. B} \textbf{\bibinfo{volume}{85}},
  \bibinfo{pages}{155440} (\bibinfo{year}{2012}).

\bibitem[{\citenamefont{Rai et~al.}(2011)\citenamefont{Rai, Hod, and
  Nitzan}}]{doi:10.1021/jz200862r}
\bibinfo{author}{\bibfnamefont{D.}~\bibnamefont{Rai}},
  \bibinfo{author}{\bibfnamefont{O.}~\bibnamefont{Hod}}, \bibnamefont{and}
  \bibinfo{author}{\bibfnamefont{A.}~\bibnamefont{Nitzan}},
  \bibinfo{journal}{The Journal of Physical Chemistry Letters}
  \textbf{\bibinfo{volume}{2}}, \bibinfo{pages}{2118} (\bibinfo{year}{2011}).

\bibitem[{\citenamefont{Solomon et~al.}(2008)\citenamefont{Solomon, Andrews,
  Hansen, Goldsmith, Wasielewski, Van~Duyne, and Ratner}}]{Solomon2008}
\bibinfo{author}{\bibfnamefont{G.~C.} \bibnamefont{Solomon}},
  \bibinfo{author}{\bibfnamefont{D.~Q.} \bibnamefont{Andrews}},
  \bibinfo{author}{\bibfnamefont{T.}~\bibnamefont{Hansen}},
  \bibinfo{author}{\bibfnamefont{R.~H.} \bibnamefont{Goldsmith}},
  \bibinfo{author}{\bibfnamefont{M.~R.} \bibnamefont{Wasielewski}},
  \bibinfo{author}{\bibfnamefont{R.~P.} \bibnamefont{Van~Duyne}},
  \bibnamefont{and} \bibinfo{author}{\bibfnamefont{M.~A.}
  \bibnamefont{Ratner}}, \bibinfo{journal}{J. Chem. Phys.}
  \textbf{\bibinfo{volume}{129}}, \bibinfo{pages}{054701}
  (\bibinfo{year}{2008}).

\bibitem[{\citenamefont{Markussen et~al.}(2010)\citenamefont{Markussen,
  Stadler, and Thygesen}}]{doi:10.1021/nl101688a}
\bibinfo{author}{\bibfnamefont{T.}~\bibnamefont{Markussen}},
  \bibinfo{author}{\bibfnamefont{R.}~\bibnamefont{Stadler}}, \bibnamefont{and}
  \bibinfo{author}{\bibfnamefont{K.~S.} \bibnamefont{Thygesen}},
  \bibinfo{journal}{Nano Letters} \textbf{\bibinfo{volume}{10}},
  \bibinfo{pages}{4260} (\bibinfo{year}{2010}).

\bibitem[{\citenamefont{Aharonov and Bohm}(1959)}]{PhysRev.115.485}
\bibinfo{author}{\bibfnamefont{Y.}~\bibnamefont{Aharonov}} \bibnamefont{and}
  \bibinfo{author}{\bibfnamefont{D.}~\bibnamefont{Bohm}},
  \bibinfo{journal}{Phys. Rev.} \textbf{\bibinfo{volume}{115}},
  \bibinfo{pages}{485} (\bibinfo{year}{1959}).

\bibitem[{\citenamefont{Aharonov and Bohm}(1961)}]{PhysRev.123.1511}
\bibinfo{author}{\bibfnamefont{Y.}~\bibnamefont{Aharonov}} \bibnamefont{and}
  \bibinfo{author}{\bibfnamefont{D.}~\bibnamefont{Bohm}},
  \bibinfo{journal}{Phys. Rev.} \textbf{\bibinfo{volume}{123}},
  \bibinfo{pages}{1511} (\bibinfo{year}{1961}).

\bibitem[{\citenamefont{Viefers et~al.}(2004)\citenamefont{Viefers, Koskinen,
  Deo, and Manninen}}]{Viefers20041}
\bibinfo{author}{\bibfnamefont{S.}~\bibnamefont{Viefers}},
  \bibinfo{author}{\bibfnamefont{P.}~\bibnamefont{Koskinen}},
  \bibinfo{author}{\bibfnamefont{P.~S.} \bibnamefont{Deo}}, \bibnamefont{and}
  \bibinfo{author}{\bibfnamefont{M.}~\bibnamefont{Manninen}},
  \bibinfo{journal}{Physica E: Low-dimensional Systems and Nanostructures}
  \textbf{\bibinfo{volume}{21}}, \bibinfo{pages}{1 } (\bibinfo{year}{2004}).

\bibitem[{\citenamefont{Bergfield et~al.}(2011)\citenamefont{Bergfield,
  Solomon, Stafford, and Ratner}}]{doi:10.1021/nl201042m}
\bibinfo{author}{\bibfnamefont{J.~P.} \bibnamefont{Bergfield}},
  \bibinfo{author}{\bibfnamefont{G.~C.} \bibnamefont{Solomon}},
  \bibinfo{author}{\bibfnamefont{C.~A.} \bibnamefont{Stafford}},
  \bibnamefont{and} \bibinfo{author}{\bibfnamefont{M.~A.}
  \bibnamefont{Ratner}}, \bibinfo{journal}{Nano Letters}
  \textbf{\bibinfo{volume}{11}}, \bibinfo{pages}{2759} (\bibinfo{year}{2011}).

\bibitem[{\citenamefont{Valli}(2013)}]{Valli2013}
\bibinfo{author}{\bibfnamefont{A.}~\bibnamefont{Valli}},
  \emph{\bibinfo{title}{Phd thesis: Electronic correlations at the nanoscale}}
  (\bibinfo{publisher}{Vienna University of Technology}, \bibinfo{year}{2013}).

\bibitem[{\citenamefont{Valli et~al.}(2012)\citenamefont{Valli, Sangiovanni,
  Toschi, and Held}}]{PhysRevB.86.115418}
\bibinfo{author}{\bibfnamefont{A.}~\bibnamefont{Valli}},
  \bibinfo{author}{\bibfnamefont{G.}~\bibnamefont{Sangiovanni}},
  \bibinfo{author}{\bibfnamefont{A.}~\bibnamefont{Toschi}}, \bibnamefont{and}
  \bibinfo{author}{\bibfnamefont{K.}~\bibnamefont{Held}},
  \bibinfo{journal}{Phys. Rev. B} \textbf{\bibinfo{volume}{86}},
  \bibinfo{pages}{115418} (\bibinfo{year}{2012}).

\bibitem[{\citenamefont{Begemann et~al.}(2008)\citenamefont{Begemann, Darau,
  Donarini, and Grifoni}}]{PhysRevB.77.201406}
\bibinfo{author}{\bibfnamefont{G.}~\bibnamefont{Begemann}},
  \bibinfo{author}{\bibfnamefont{D.}~\bibnamefont{Darau}},
  \bibinfo{author}{\bibfnamefont{A.}~\bibnamefont{Donarini}}, \bibnamefont{and}
  \bibinfo{author}{\bibfnamefont{M.}~\bibnamefont{Grifoni}},
  \bibinfo{journal}{Phys. Rev. B} \textbf{\bibinfo{volume}{77}},
  \bibinfo{pages}{201406} (\bibinfo{year}{2008}).

\bibitem[{\citenamefont{Darau et~al.}(2009)\citenamefont{Darau, Begemann,
  Donarini, and Grifoni}}]{PhysRevB.79.235404}
\bibinfo{author}{\bibfnamefont{D.}~\bibnamefont{Darau}},
  \bibinfo{author}{\bibfnamefont{G.}~\bibnamefont{Begemann}},
  \bibinfo{author}{\bibfnamefont{A.}~\bibnamefont{Donarini}}, \bibnamefont{and}
  \bibinfo{author}{\bibfnamefont{M.}~\bibnamefont{Grifoni}},
  \bibinfo{journal}{Phys. Rev. B} \textbf{\bibinfo{volume}{79}},
  \bibinfo{pages}{235404} (\bibinfo{year}{2009}).

\bibitem[{\citenamefont{Bergfield and
  Stafford}(2009{\natexlab{a}})}]{PhysRevB.79.245125}
\bibinfo{author}{\bibfnamefont{J.~P.} \bibnamefont{Bergfield}}
  \bibnamefont{and} \bibinfo{author}{\bibfnamefont{C.~A.}
  \bibnamefont{Stafford}}, \bibinfo{journal}{Phys. Rev. B}
  \textbf{\bibinfo{volume}{79}}, \bibinfo{pages}{245125}
  (\bibinfo{year}{2009}{\natexlab{a}}).

\bibitem[{\citenamefont{Bergfield and
  Stafford}(2009{\natexlab{b}})}]{doi:10.1021/nl901554s}
\bibinfo{author}{\bibfnamefont{J.~P.} \bibnamefont{Bergfield}}
  \bibnamefont{and} \bibinfo{author}{\bibfnamefont{C.~A.}
  \bibnamefont{Stafford}}, \bibinfo{journal}{Nano Letters}
  \textbf{\bibinfo{volume}{9}}, \bibinfo{pages}{3072}
  (\bibinfo{year}{2009}{\natexlab{b}}).

\bibitem[{\citenamefont{Bohr and
  Schmitteckert}(2012{\natexlab{a}})}]{ANDP:ANDP201100266}
\bibinfo{author}{\bibfnamefont{D.}~\bibnamefont{Bohr}} \bibnamefont{and}
  \bibinfo{author}{\bibfnamefont{P.}~\bibnamefont{Schmitteckert}},
  \bibinfo{journal}{Annalen der Physik} \textbf{\bibinfo{volume}{524}},
  \bibinfo{pages}{199} (\bibinfo{year}{2012}{\natexlab{a}}).

\bibitem[{\citenamefont{Leijnse et~al.}(2011)\citenamefont{Leijnse, Sun,
  Brondsted~Nielsen, Hedegard, and Flensberg}}]{Leijnse_2011}
\bibinfo{author}{\bibfnamefont{M.}~\bibnamefont{Leijnse}},
  \bibinfo{author}{\bibfnamefont{W.}~\bibnamefont{Sun}},
  \bibinfo{author}{\bibfnamefont{M.}~\bibnamefont{Brondsted~Nielsen}},
  \bibinfo{author}{\bibfnamefont{P.}~\bibnamefont{Hedegard}}, \bibnamefont{and}
  \bibinfo{author}{\bibfnamefont{K.}~\bibnamefont{Flensberg}},
  \bibinfo{journal}{J. Chem. Phys.} \textbf{\bibinfo{volume}{134}},
  \bibinfo{pages}{104107} (\bibinfo{year}{2011}).

\bibitem[{\citenamefont{Ryndyk et~al.}(2012)\citenamefont{Ryndyk, Bundesmann,
  Liu, and Richter}}]{PhysRevB.86.195425}
\bibinfo{author}{\bibfnamefont{D.~A.} \bibnamefont{Ryndyk}},
  \bibinfo{author}{\bibfnamefont{J.}~\bibnamefont{Bundesmann}},
  \bibinfo{author}{\bibfnamefont{M.-H.} \bibnamefont{Liu}}, \bibnamefont{and}
  \bibinfo{author}{\bibfnamefont{K.}~\bibnamefont{Richter}},
  \bibinfo{journal}{Phys. Rev. B} \textbf{\bibinfo{volume}{86}},
  \bibinfo{pages}{195425} (\bibinfo{year}{2012}).

\bibitem[{\citenamefont{Hewson}(1997)}]{hewson_kondo_1997}
\bibinfo{author}{\bibfnamefont{A.~C.} \bibnamefont{Hewson}},
  \emph{\bibinfo{title}{The Kondo Problem to Heavy Fermions}}
  (\bibinfo{publisher}{Cambridge University Press}, \bibinfo{year}{1997}), ISBN
  \bibinfo{isbn}{0521599474}.

\bibitem[{\citenamefont{Bohr and
  Schmitteckert}(2012{\natexlab{b}})}]{Bohr_2012}
\bibinfo{author}{\bibfnamefont{D.}~\bibnamefont{Bohr}} \bibnamefont{and}
  \bibinfo{author}{\bibfnamefont{P.}~\bibnamefont{Schmitteckert}},
  \bibinfo{journal}{Ann. Phys.} \textbf{\bibinfo{volume}{524}},
  \bibinfo{pages}{199204} (\bibinfo{year}{2012}{\natexlab{b}}).

\bibitem[{\citenamefont{Yu et~al.}(2005)\citenamefont{Yu, Keane, Ciszek, Cheng,
  Tour, Baruah, Pederson, and Natelson}}]{PhysRevLett.95.256803}
\bibinfo{author}{\bibfnamefont{L.~H.} \bibnamefont{Yu}},
  \bibinfo{author}{\bibfnamefont{Z.~K.} \bibnamefont{Keane}},
  \bibinfo{author}{\bibfnamefont{J.~W.} \bibnamefont{Ciszek}},
  \bibinfo{author}{\bibfnamefont{L.}~\bibnamefont{Cheng}},
  \bibinfo{author}{\bibfnamefont{J.~M.} \bibnamefont{Tour}},
  \bibinfo{author}{\bibfnamefont{T.}~\bibnamefont{Baruah}},
  \bibinfo{author}{\bibfnamefont{M.~R.} \bibnamefont{Pederson}},
  \bibnamefont{and} \bibinfo{author}{\bibfnamefont{D.}~\bibnamefont{Natelson}},
  \bibinfo{journal}{Phys. Rev. Lett.} \textbf{\bibinfo{volume}{95}},
  \bibinfo{pages}{256803} (\bibinfo{year}{2005}).

\bibitem[{\citenamefont{Tosi et~al.}(2012)\citenamefont{Tosi, Roura-Bas, and
  Aligia}}]{Tosi2012}
\bibinfo{author}{\bibfnamefont{L.}~\bibnamefont{Tosi}},
  \bibinfo{author}{\bibfnamefont{P.}~\bibnamefont{Roura-Bas}},
  \bibnamefont{and} \bibinfo{author}{\bibfnamefont{A.~A.}
  \bibnamefont{Aligia}}, \bibinfo{journal}{Journal of Physics: Condensed
  Matter} \textbf{\bibinfo{volume}{24}}, \bibinfo{pages}{365301}
  (\bibinfo{year}{2012}).

\bibitem[{\citenamefont{Nichols et~al.}(2010)\citenamefont{Nichols, Haiss,
  Higgins, Leary, Martin, and Bethell}}]{B922000C}
\bibinfo{author}{\bibfnamefont{R.~J.} \bibnamefont{Nichols}},
  \bibinfo{author}{\bibfnamefont{W.}~\bibnamefont{Haiss}},
  \bibinfo{author}{\bibfnamefont{S.~J.} \bibnamefont{Higgins}},
  \bibinfo{author}{\bibfnamefont{E.}~\bibnamefont{Leary}},
  \bibinfo{author}{\bibfnamefont{S.}~\bibnamefont{Martin}}, \bibnamefont{and}
  \bibinfo{author}{\bibfnamefont{D.}~\bibnamefont{Bethell}},
  \bibinfo{journal}{Phys. Chem. Chem. Phys.} \textbf{\bibinfo{volume}{12}},
  \bibinfo{pages}{2801} (\bibinfo{year}{2010}).

\bibitem[{\citenamefont{Datta}(2005)}]{Datta_2005}
\bibinfo{author}{\bibfnamefont{S.}~\bibnamefont{Datta}},
  \emph{\bibinfo{title}{Quantum Transport: Atom to Transistor}}
  (\bibinfo{publisher}{Cambridge University Press}, \bibinfo{year}{2005}),
  \bibinfo{edition}{2nd} ed., ISBN \bibinfo{isbn}{0521631459}.

\bibitem[{\citenamefont{Ventra}(2008)}]{Ventra_2008}
\bibinfo{author}{\bibfnamefont{M.~D.} \bibnamefont{Ventra}},
  \emph{\bibinfo{title}{Electrical Transport in Nanoscale Systems}}
  (\bibinfo{publisher}{Cambridge University Press, New York},
  \bibinfo{year}{2008}), ISBN \bibinfo{isbn}{0521896347}.

\bibitem[{\citenamefont{Ferry et~al.}(2009)\citenamefont{Ferry, Goodnick, and
  Bird}}]{Ferry_2009}
\bibinfo{author}{\bibfnamefont{D.~K.} \bibnamefont{Ferry}},
  \bibinfo{author}{\bibfnamefont{S.~M.} \bibnamefont{Goodnick}},
  \bibnamefont{and} \bibinfo{author}{\bibfnamefont{J.}~\bibnamefont{Bird}},
  \emph{\bibinfo{title}{Transport in Nanostructures}}
  (\bibinfo{publisher}{Cambridge University Press}, \bibinfo{year}{2009}),
  \bibinfo{edition}{2nd} ed., ISBN \bibinfo{isbn}{0521877482}.

\bibitem[{\citenamefont{Nazarov and Blanter}(2009)}]{Nazarov_2009}
\bibinfo{author}{\bibfnamefont{Y.~V.} \bibnamefont{Nazarov}} \bibnamefont{and}
  \bibinfo{author}{\bibfnamefont{Y.~M.} \bibnamefont{Blanter}},
  \emph{\bibinfo{title}{Quantum Transport: Introduction to Nanoscience}}
  (\bibinfo{publisher}{Cambridge University Press New York},
  \bibinfo{year}{2009}), ISBN \bibinfo{isbn}{0521832462}.

\bibitem[{\citenamefont{Venkataraman
  et~al.}(2006{\natexlab{b}})\citenamefont{Venkataraman, Klare, Tam, Nuckolls,
  Hybertsen, and Steigerwald}}]{doi:10.1021/nl052373+}
\bibinfo{author}{\bibfnamefont{L.}~\bibnamefont{Venkataraman}},
  \bibinfo{author}{\bibfnamefont{J.~E.} \bibnamefont{Klare}},
  \bibinfo{author}{\bibfnamefont{I.~W.} \bibnamefont{Tam}},
  \bibinfo{author}{\bibfnamefont{C.}~\bibnamefont{Nuckolls}},
  \bibinfo{author}{\bibfnamefont{M.~S.} \bibnamefont{Hybertsen}},
  \bibnamefont{and} \bibinfo{author}{\bibfnamefont{M.~L.}
  \bibnamefont{Steigerwald}}, \bibinfo{journal}{Nano Letters}
  \textbf{\bibinfo{volume}{6}}, \bibinfo{pages}{458}
  (\bibinfo{year}{2006}{\natexlab{b}}).

\bibitem[{\citenamefont{Richter}(1999)}]{richter_semiclassical_1999}
\bibinfo{author}{\bibfnamefont{K.}~\bibnamefont{Richter}},
  \emph{\bibinfo{title}{Semiclassical Theory of Mesoscopic Quantum Systems}}
  (\bibinfo{publisher}{Springer}, \bibinfo{year}{1999}), ISBN
  \bibinfo{isbn}{3540665668}.

\bibitem[{\citenamefont{Caroli et~al.}(1971)\citenamefont{Caroli, Combescot,
  Nozieres, and Saint-James}}]{0022-3719-4-8-018}
\bibinfo{author}{\bibfnamefont{C.}~\bibnamefont{Caroli}},
  \bibinfo{author}{\bibfnamefont{R.}~\bibnamefont{Combescot}},
  \bibinfo{author}{\bibfnamefont{P.}~\bibnamefont{Nozieres}}, \bibnamefont{and}
  \bibinfo{author}{\bibfnamefont{D.}~\bibnamefont{Saint-James}},
  \bibinfo{journal}{Journal of Physics C: Solid State Physics}
  \textbf{\bibinfo{volume}{4}}, \bibinfo{pages}{916} (\bibinfo{year}{1971}).

\bibitem[{\citenamefont{Hohenberg and Kohn}(1964)}]{PhysRev.136.B864}
\bibinfo{author}{\bibfnamefont{P.}~\bibnamefont{Hohenberg}} \bibnamefont{and}
  \bibinfo{author}{\bibfnamefont{W.}~\bibnamefont{Kohn}},
  \bibinfo{journal}{Phys. Rev.} \textbf{\bibinfo{volume}{136}},
  \bibinfo{pages}{B864} (\bibinfo{year}{1964}).

\bibitem[{\citenamefont{Kohn and Sham}(1965)}]{PhysRev.140.A1133}
\bibinfo{author}{\bibfnamefont{W.}~\bibnamefont{Kohn}} \bibnamefont{and}
  \bibinfo{author}{\bibfnamefont{L.~J.} \bibnamefont{Sham}},
  \bibinfo{journal}{Phys. Rev.} \textbf{\bibinfo{volume}{140}},
  \bibinfo{pages}{A1133} (\bibinfo{year}{1965}).

\bibitem[{\citenamefont{Frederiksen et~al.}(2007)\citenamefont{Frederiksen,
  Paulsson, Brandbyge, and Jauho}}]{PhysRevB.75.205413}
\bibinfo{author}{\bibfnamefont{T.}~\bibnamefont{Frederiksen}},
  \bibinfo{author}{\bibfnamefont{M.}~\bibnamefont{Paulsson}},
  \bibinfo{author}{\bibfnamefont{M.}~\bibnamefont{Brandbyge}},
  \bibnamefont{and} \bibinfo{author}{\bibfnamefont{A.-P.} \bibnamefont{Jauho}},
  \bibinfo{journal}{Phys. Rev. B} \textbf{\bibinfo{volume}{75}},
  \bibinfo{pages}{205413} (\bibinfo{year}{2007}).

\bibitem[{\citenamefont{Taylor et~al.}(2001)\citenamefont{Taylor, Guo, and
  Wang}}]{PhysRevB.63.245407}
\bibinfo{author}{\bibfnamefont{J.}~\bibnamefont{Taylor}},
  \bibinfo{author}{\bibfnamefont{H.}~\bibnamefont{Guo}}, \bibnamefont{and}
  \bibinfo{author}{\bibfnamefont{J.}~\bibnamefont{Wang}},
  \bibinfo{journal}{Phys. Rev. B} \textbf{\bibinfo{volume}{63}},
  \bibinfo{pages}{245407} (\bibinfo{year}{2001}).

\bibitem[{\citenamefont{Brandbyge et~al.}(2002)\citenamefont{Brandbyge, Mozos,
  Ordej\'on, Taylor, and Stokbro}}]{PhysRevB.65.165401}
\bibinfo{author}{\bibfnamefont{M.}~\bibnamefont{Brandbyge}},
  \bibinfo{author}{\bibfnamefont{J.-L.} \bibnamefont{Mozos}},
  \bibinfo{author}{\bibfnamefont{P.}~\bibnamefont{Ordej\'on}},
  \bibinfo{author}{\bibfnamefont{J.}~\bibnamefont{Taylor}}, \bibnamefont{and}
  \bibinfo{author}{\bibfnamefont{K.}~\bibnamefont{Stokbro}},
  \bibinfo{journal}{Phys. Rev. B} \textbf{\bibinfo{volume}{65}},
  \bibinfo{pages}{165401} (\bibinfo{year}{2002}).

\bibitem[{\citenamefont{Fujimoto and Hirose}(2003)}]{PhysRevB.67.195315}
\bibinfo{author}{\bibfnamefont{Y.}~\bibnamefont{Fujimoto}} \bibnamefont{and}
  \bibinfo{author}{\bibfnamefont{K.}~\bibnamefont{Hirose}},
  \bibinfo{journal}{Phys. Rev. B} \textbf{\bibinfo{volume}{67}},
  \bibinfo{pages}{195315} (\bibinfo{year}{2003}).

\bibitem[{\citenamefont{Rocha et~al.}(2006)\citenamefont{Rocha, Garcia-Suarez,
  Bailey, Lambert, Ferrer, and Sanvito}}]{PhysRevB.73.085414}
\bibinfo{author}{\bibfnamefont{A.~R.} \bibnamefont{Rocha}},
  \bibinfo{author}{\bibfnamefont{V.~M.} \bibnamefont{Garcia-Suarez}},
  \bibinfo{author}{\bibfnamefont{S.}~\bibnamefont{Bailey}},
  \bibinfo{author}{\bibfnamefont{C.}~\bibnamefont{Lambert}},
  \bibinfo{author}{\bibfnamefont{J.}~\bibnamefont{Ferrer}}, \bibnamefont{and}
  \bibinfo{author}{\bibfnamefont{S.}~\bibnamefont{Sanvito}},
  \bibinfo{journal}{Phys. Rev. B} \textbf{\bibinfo{volume}{73}},
  \bibinfo{pages}{085414} (\bibinfo{year}{2006}).

\bibitem[{\citenamefont{Calzolari et~al.}(2004)\citenamefont{Calzolari,
  Marzari, Souza, and Buongiorno~Nardelli}}]{PhysRevB.69.035108}
\bibinfo{author}{\bibfnamefont{A.}~\bibnamefont{Calzolari}},
  \bibinfo{author}{\bibfnamefont{N.}~\bibnamefont{Marzari}},
  \bibinfo{author}{\bibfnamefont{I.}~\bibnamefont{Souza}}, \bibnamefont{and}
  \bibinfo{author}{\bibfnamefont{M.}~\bibnamefont{Buongiorno~Nardelli}},
  \bibinfo{journal}{Phys. Rev. B} \textbf{\bibinfo{volume}{69}},
  \bibinfo{pages}{035108} (\bibinfo{year}{2004}).

\bibitem[{\citenamefont{Thygesen and Jacobsen}(2005)}]{Thygesen2005111}
\bibinfo{author}{\bibfnamefont{K.}~\bibnamefont{Thygesen}} \bibnamefont{and}
  \bibinfo{author}{\bibfnamefont{K.}~\bibnamefont{Jacobsen}},
  \bibinfo{journal}{Chemical Physics} \textbf{\bibinfo{volume}{319}},
  \bibinfo{pages}{111 } (\bibinfo{year}{2005}).

\bibitem[{\citenamefont{Li and Kosov}(2006)}]{0953-8984-18-4-019}
\bibinfo{author}{\bibfnamefont{Z.}~\bibnamefont{Li}} \bibnamefont{and}
  \bibinfo{author}{\bibfnamefont{D.~S.} \bibnamefont{Kosov}},
  \bibinfo{journal}{Journal of Physics: Condensed Matter}
  \textbf{\bibinfo{volume}{18}}, \bibinfo{pages}{1347} (\bibinfo{year}{2006}).

\bibitem[{\citenamefont{Perrine et~al.}(2008)\citenamefont{Perrine, Berto, and
  Dunietz}}]{doi:10.1021/jp8075854}
\bibinfo{author}{\bibfnamefont{T.~M.} \bibnamefont{Perrine}},
  \bibinfo{author}{\bibfnamefont{T.}~\bibnamefont{Berto}}, \bibnamefont{and}
  \bibinfo{author}{\bibfnamefont{B.~D.} \bibnamefont{Dunietz}},
  \bibinfo{journal}{The Journal of Physical Chemistry B}
  \textbf{\bibinfo{volume}{112}}, \bibinfo{pages}{16070}
  (\bibinfo{year}{2008}).

\bibitem[{\citenamefont{Arnold et~al.}(2007)\citenamefont{Arnold, Weigend, and
  Evers}}]{Arnold_2007}
\bibinfo{author}{\bibfnamefont{A.}~\bibnamefont{Arnold}},
  \bibinfo{author}{\bibfnamefont{F.}~\bibnamefont{Weigend}}, \bibnamefont{and}
  \bibinfo{author}{\bibfnamefont{F.}~\bibnamefont{Evers}}, \bibinfo{journal}{J.
  Chem. Phys.} \textbf{\bibinfo{volume}{126}}, \bibinfo{pages}{174101}
  (\bibinfo{year}{2007}).

\bibitem[{foo({\natexlab{a}})}]{footnote1}
\bibinfo{note}{Recently it was found that using the DFT auxiliary (Kohn-Sham)
  particles in Kubo linear response theory, the conductance through a benzene
  molecular junction is zero for all bias voltages, while a "non-interacting"
  description in terms of the reduced density matrix leads to conduction on the
  correct order of magnitude.~\cite{Schmitteckert2013}}.

\bibitem[{\citenamefont{Dzhioev and Kosov}(2011)}]{Dzhioev_2011}
\bibinfo{author}{\bibfnamefont{A.~A.} \bibnamefont{Dzhioev}} \bibnamefont{and}
  \bibinfo{author}{\bibfnamefont{D.~S.} \bibnamefont{Kosov}},
  \bibinfo{journal}{J. Chem. Phys.} \textbf{\bibinfo{volume}{134}},
  \bibinfo{pages}{044121} (\bibinfo{year}{2011}).

\bibitem[{\citenamefont{Toher et~al.}(2005)\citenamefont{Toher, Filippetti,
  Sanvito, and Burke}}]{PhysRevLett.95.146402}
\bibinfo{author}{\bibfnamefont{C.}~\bibnamefont{Toher}},
  \bibinfo{author}{\bibfnamefont{A.}~\bibnamefont{Filippetti}},
  \bibinfo{author}{\bibfnamefont{S.}~\bibnamefont{Sanvito}}, \bibnamefont{and}
  \bibinfo{author}{\bibfnamefont{K.}~\bibnamefont{Burke}},
  \bibinfo{journal}{Phys. Rev. Lett.} \textbf{\bibinfo{volume}{95}},
  \bibinfo{pages}{146402} (\bibinfo{year}{2005}).

\bibitem[{\citenamefont{Delaney and Greer}(2004)}]{PhysRevLett.93.036805}
\bibinfo{author}{\bibfnamefont{P.}~\bibnamefont{Delaney}} \bibnamefont{and}
  \bibinfo{author}{\bibfnamefont{J.~C.} \bibnamefont{Greer}},
  \bibinfo{journal}{Phys. Rev. Lett.} \textbf{\bibinfo{volume}{93}},
  \bibinfo{pages}{036805} (\bibinfo{year}{2004}).

\bibitem[{\citenamefont{Strange et~al.}(2008)\citenamefont{Strange, Kristensen,
  Thygesen, and Jacobsen}}]{Strange2008}
\bibinfo{author}{\bibfnamefont{M.}~\bibnamefont{Strange}},
  \bibinfo{author}{\bibfnamefont{I.~S.} \bibnamefont{Kristensen}},
  \bibinfo{author}{\bibfnamefont{K.~S.} \bibnamefont{Thygesen}},
  \bibnamefont{and} \bibinfo{author}{\bibfnamefont{K.~W.}
  \bibnamefont{Jacobsen}}, \bibinfo{journal}{J. Chem. Phys.}
  \textbf{\bibinfo{volume}{128}}, \bibinfo{pages}{114714}
  (\bibinfo{year}{2008}).

\bibitem[{\citenamefont{Lee et~al.}(2009)\citenamefont{Lee, Jean, and
  Sanvito}}]{PhysRevB.79.085120}
\bibinfo{author}{\bibfnamefont{W.}~\bibnamefont{Lee}},
  \bibinfo{author}{\bibfnamefont{N.}~\bibnamefont{Jean}}, \bibnamefont{and}
  \bibinfo{author}{\bibfnamefont{S.}~\bibnamefont{Sanvito}},
  \bibinfo{journal}{Phys. Rev. B} \textbf{\bibinfo{volume}{79}},
  \bibinfo{pages}{085120} (\bibinfo{year}{2009}).

\bibitem[{\citenamefont{Baer and Neuhauser}(2003)}]{Baer2003459}
\bibinfo{author}{\bibfnamefont{R.}~\bibnamefont{Baer}} \bibnamefont{and}
  \bibinfo{author}{\bibfnamefont{D.}~\bibnamefont{Neuhauser}},
  \bibinfo{journal}{Chemical Physics Letters} \textbf{\bibinfo{volume}{374}},
  \bibinfo{pages}{459 } (\bibinfo{year}{2003}).

\bibitem[{\citenamefont{Thygesen and Rubio}(2008)}]{PhysRevB.77.115333}
\bibinfo{author}{\bibfnamefont{K.~S.} \bibnamefont{Thygesen}} \bibnamefont{and}
  \bibinfo{author}{\bibfnamefont{A.}~\bibnamefont{Rubio}},
  \bibinfo{journal}{Phys. Rev. B} \textbf{\bibinfo{volume}{77}},
  \bibinfo{pages}{115333} (\bibinfo{year}{2008}).

\bibitem[{\citenamefont{My\"oh\"anen et~al.}(2009)\citenamefont{My\"oh\"anen,
  Stan, Stefanucci, and van Leeuwen}}]{PhysRevB.80.115107}
\bibinfo{author}{\bibfnamefont{P.}~\bibnamefont{My\"oh\"anen}},
  \bibinfo{author}{\bibfnamefont{A.}~\bibnamefont{Stan}},
  \bibinfo{author}{\bibfnamefont{G.}~\bibnamefont{Stefanucci}},
  \bibnamefont{and} \bibinfo{author}{\bibfnamefont{R.}~\bibnamefont{van
  Leeuwen}}, \bibinfo{journal}{Phys. Rev. B} \textbf{\bibinfo{volume}{80}},
  \bibinfo{pages}{115107} (\bibinfo{year}{2009}).

\bibitem[{\citenamefont{Spataru et~al.}(2009)\citenamefont{Spataru, Hybertsen,
  Louie, and Millis}}]{PhysRevB.79.155110}
\bibinfo{author}{\bibfnamefont{C.~D.} \bibnamefont{Spataru}},
  \bibinfo{author}{\bibfnamefont{M.~S.} \bibnamefont{Hybertsen}},
  \bibinfo{author}{\bibfnamefont{S.~G.} \bibnamefont{Louie}}, \bibnamefont{and}
  \bibinfo{author}{\bibfnamefont{A.~J.} \bibnamefont{Millis}},
  \bibinfo{journal}{Phys. Rev. B} \textbf{\bibinfo{volume}{79}},
  \bibinfo{pages}{155110} (\bibinfo{year}{2009}).

\bibitem[{\citenamefont{Dahnovsky}(2009)}]{PhysRevB.80.165305}
\bibinfo{author}{\bibfnamefont{Y.}~\bibnamefont{Dahnovsky}},
  \bibinfo{journal}{Phys. Rev. B} \textbf{\bibinfo{volume}{80}},
  \bibinfo{pages}{165305} (\bibinfo{year}{2009}).

\bibitem[{\citenamefont{Arrigoni et~al.}(2013)\citenamefont{Arrigoni, Knap, and
  von~der Linden}}]{Arrigoni2012}
\bibinfo{author}{\bibfnamefont{E.}~\bibnamefont{Arrigoni}},
  \bibinfo{author}{\bibfnamefont{M.}~\bibnamefont{Knap}}, \bibnamefont{and}
  \bibinfo{author}{\bibfnamefont{W.}~\bibnamefont{von~der Linden}},
  \bibinfo{journal}{Phys. Rev. Lett.} \textbf{\bibinfo{volume}{110}},
  \bibinfo{pages}{086403} (\bibinfo{year}{2013}).

\bibitem[{\citenamefont{Bursill et~al.}(1998)\citenamefont{Bursill, Castleton,
  and Barford}}]{Bursill1998305}
\bibinfo{author}{\bibfnamefont{R.~J.} \bibnamefont{Bursill}},
  \bibinfo{author}{\bibfnamefont{C.}~\bibnamefont{Castleton}},
  \bibnamefont{and} \bibinfo{author}{\bibfnamefont{W.}~\bibnamefont{Barford}},
  \bibinfo{journal}{Chemical Physics Letters} \textbf{\bibinfo{volume}{294}},
  \bibinfo{pages}{305 } (\bibinfo{year}{1998}).

\bibitem[{\citenamefont{William}(2005)}]{barford_2005}
\bibinfo{author}{\bibfnamefont{B.}~\bibnamefont{William}},
  \emph{\bibinfo{title}{Electronic and Optical Properties of Conjugated
  Polymers}} (\bibinfo{publisher}{Oxford University Press},
  \bibinfo{year}{2005}), ISBN \bibinfo{isbn}{0198526806}.

\bibitem[{\citenamefont{Hettler et~al.}(2003)\citenamefont{Hettler, Wenzel,
  Wegewijs, and Schoeller}}]{PhysRevLett.90.076805}
\bibinfo{author}{\bibfnamefont{M.~H.} \bibnamefont{Hettler}},
  \bibinfo{author}{\bibfnamefont{W.}~\bibnamefont{Wenzel}},
  \bibinfo{author}{\bibfnamefont{M.~R.} \bibnamefont{Wegewijs}},
  \bibnamefont{and}
  \bibinfo{author}{\bibfnamefont{H.}~\bibnamefont{Schoeller}},
  \bibinfo{journal}{Phys. Rev. Lett.} \textbf{\bibinfo{volume}{90}},
  \bibinfo{pages}{076805} (\bibinfo{year}{2003}).

\bibitem[{\citenamefont{Ryndyk et~al.}(2013)\citenamefont{Ryndyk, Donarini,
  Grifoni, and Richter}}]{PhysRevB.88.085404}
\bibinfo{author}{\bibfnamefont{D.~A.} \bibnamefont{Ryndyk}},
  \bibinfo{author}{\bibfnamefont{A.}~\bibnamefont{Donarini}},
  \bibinfo{author}{\bibfnamefont{M.}~\bibnamefont{Grifoni}}, \bibnamefont{and}
  \bibinfo{author}{\bibfnamefont{K.}~\bibnamefont{Richter}},
  \bibinfo{journal}{Phys. Rev. B} \textbf{\bibinfo{volume}{88}},
  \bibinfo{pages}{085404} (\bibinfo{year}{2013}).

\bibitem[{\citenamefont{Kaasbjerg and Flensberg}(2008)}]{doi:10.1021/nl8021708}
\bibinfo{author}{\bibfnamefont{K.}~\bibnamefont{Kaasbjerg}} \bibnamefont{and}
  \bibinfo{author}{\bibfnamefont{K.}~\bibnamefont{Flensberg}},
  \bibinfo{journal}{Nano Letters} \textbf{\bibinfo{volume}{8}},
  \bibinfo{pages}{3809} (\bibinfo{year}{2008}).

\bibitem[{\citenamefont{Kubatkin et~al.}(2003)\citenamefont{Kubatkin, Danilov,
  Hjort, Cornil, Bredas, Stuhr-Hansen, Hedegard, and
  Bjornholm}}]{kubatkin_single-electron_2003}
\bibinfo{author}{\bibfnamefont{S.}~\bibnamefont{Kubatkin}},
  \bibinfo{author}{\bibfnamefont{A.}~\bibnamefont{Danilov}},
  \bibinfo{author}{\bibfnamefont{M.}~\bibnamefont{Hjort}},
  \bibinfo{author}{\bibfnamefont{J.}~\bibnamefont{Cornil}},
  \bibinfo{author}{\bibfnamefont{J.-L.} \bibnamefont{Bredas}},
  \bibinfo{author}{\bibfnamefont{N.}~\bibnamefont{Stuhr-Hansen}},
  \bibinfo{author}{\bibfnamefont{P.}~\bibnamefont{Hedegard}}, \bibnamefont{and}
  \bibinfo{author}{\bibfnamefont{T.}~\bibnamefont{Bjornholm}},
  \bibinfo{journal}{Nature} \textbf{\bibinfo{volume}{425}},
  \bibinfo{pages}{698} (\bibinfo{year}{2003}).

\bibitem[{\citenamefont{Pariser and Parr}(1953)}]{Pariser_1953}
\bibinfo{author}{\bibfnamefont{R.}~\bibnamefont{Pariser}} \bibnamefont{and}
  \bibinfo{author}{\bibfnamefont{R.}~\bibnamefont{Parr}},
  \bibinfo{journal}{Journal of Chemical Physics} \textbf{\bibinfo{volume}{21}},
  \bibinfo{pages}{767} (\bibinfo{year}{1953}).

\bibitem[{\citenamefont{Pople}(1953)}]{Pople_1953}
\bibinfo{author}{\bibfnamefont{J.~A.} \bibnamefont{Pople}},
  \bibinfo{journal}{Transactions of the Faraday Society}
  \textbf{\bibinfo{volume}{49}}, \bibinfo{pages}{1375} (\bibinfo{year}{1953}).

\bibitem[{\citenamefont{Balzer and
  Potthoff}(2011)}]{balzer_non-equilibrium_2011}
\bibinfo{author}{\bibfnamefont{M.}~\bibnamefont{Balzer}} \bibnamefont{and}
  \bibinfo{author}{\bibfnamefont{M.}~\bibnamefont{Potthoff}},
  \bibinfo{journal}{Phys. Rev. B} \textbf{\bibinfo{volume}{83}},
  \bibinfo{pages}{195132} (\bibinfo{year}{2011}).

\bibitem[{\citenamefont{Knap et~al.}(2011)\citenamefont{Knap, von~der Linden,
  and Arrigoni}}]{knap_nonequilibrium_2011}
\bibinfo{author}{\bibfnamefont{M.}~\bibnamefont{Knap}},
  \bibinfo{author}{\bibfnamefont{W.}~\bibnamefont{von~der Linden}},
  \bibnamefont{and} \bibinfo{author}{\bibfnamefont{E.}~\bibnamefont{Arrigoni}},
  \bibinfo{journal}{Phys. Rev. B} \textbf{\bibinfo{volume}{84}},
  \bibinfo{pages}{115145} (\bibinfo{year}{2011}).

\bibitem[{\citenamefont{Nuss et~al.}(2012{\natexlab{a}})\citenamefont{Nuss,
  Heil, Ganahl, Knap, Evertz, Arrigoni, and von~der
  Linden}}]{PhysRevB.86.245119}
\bibinfo{author}{\bibfnamefont{M.}~\bibnamefont{Nuss}},
  \bibinfo{author}{\bibfnamefont{C.}~\bibnamefont{Heil}},
  \bibinfo{author}{\bibfnamefont{M.}~\bibnamefont{Ganahl}},
  \bibinfo{author}{\bibfnamefont{M.}~\bibnamefont{Knap}},
  \bibinfo{author}{\bibfnamefont{H.~G.} \bibnamefont{Evertz}},
  \bibinfo{author}{\bibfnamefont{E.}~\bibnamefont{Arrigoni}}, \bibnamefont{and}
  \bibinfo{author}{\bibfnamefont{W.}~\bibnamefont{von~der Linden}},
  \bibinfo{journal}{Phys. Rev. B} \textbf{\bibinfo{volume}{86}},
  \bibinfo{pages}{245119} (\bibinfo{year}{2012}{\natexlab{a}}).

\bibitem[{\citenamefont{Hubbard}(1963)}]{Hubbard_1963}
\bibinfo{author}{\bibfnamefont{J.}~\bibnamefont{Hubbard}},
  \bibinfo{journal}{Proc. R. Soc. Lond. A} \textbf{\bibinfo{volume}{276}},
  \bibinfo{pages}{238} (\bibinfo{year}{1963}).

\bibitem[{\citenamefont{Rai et~al.}(2010)\citenamefont{Rai, Hod, and
  Nitzan}}]{doi:10.1021/jp105030d}
\bibinfo{author}{\bibfnamefont{D.}~\bibnamefont{Rai}},
  \bibinfo{author}{\bibfnamefont{O.}~\bibnamefont{Hod}}, \bibnamefont{and}
  \bibinfo{author}{\bibfnamefont{A.}~\bibnamefont{Nitzan}},
  \bibinfo{journal}{The Journal of Physical Chemistry C}
  \textbf{\bibinfo{volume}{114}}, \bibinfo{pages}{20583}
  (\bibinfo{year}{2010}).

\bibitem[{NIS(2012)}]{NIST}
\emph{\bibinfo{title}{National institute for standards and technology,
  computational chemistry comparison and benchmark data base}},
  \bibinfo{howpublished}{\url{http://cccbdb.nist.gov/}} (\bibinfo{year}{2012}),
  \bibinfo{note}{[Online; accessed 28-December-2012]}.

\bibitem[{foo({\natexlab{b}})}]{footnote2}
\bibinfo{note}{Bursill \etal~\cite{Bursill1998305} used an Ohno parametrization
  for off diagonal interaction parameters which in our notation reads
  $W_{ij}[eV]=\frac{U[eV]}{2\sqrt{1+\frac{U[eV]r_{ij}[\text{\AA}]}{14.397}}}$
  and found $U\approx10.06\,eV$. In literature often only nearest-neighbor
  density-density interactions are included using
  $W=6\,eV$.~\cite{PhysRevB.77.201406,PhysRevB.79.235404} Very recently it was
  shown that for the model under discussion it is possible to take non-local
  Coulomb interaction into account in an effective model with reduced on-site
  Coulomb interaction only.~\cite{Schuler2013} Based on a constrained Random
  Phase Approximation calculation they found a value for the bare on-site
  interaction $\frac{U}{t}=3.96$ which reduces in the model with on-site
  interactions only by more than a factor of two: $\frac{U^*}{t}=1.2$.}

\bibitem[{foo({\natexlab{c}})}]{footnote11}
\bibinfo{note}{We carry out the Peierls substitution in a standard
  way~\cite{PhysRevB.85.155440} for fields $B$ perpendicular to the ring plane.
  The geometry of the ring enters through the expression $(\mathbf{r}_i\times
  \mathbf{r}_j)\cdot\hat{e}_z$ which is a parallelepiped consisting of one
  third of the area of the full ring. Using the lattice constant of benzene
  $a=1.39\,\text{\AA}$~\cite{NIST} one obtains nearest-neighbor hopping in the
  ring structure of $\tilde{t}_{d}=t_d\,e^{i
  \frac{\sqrt{3}a^2}{4}\frac{B[T]}{2.35.10^5}}=t_d\,e^{i
  1.2713.10^{-5}B[T]}=t_d\,e^{i \Phi(B)}$.}

\bibitem[{foo({\natexlab{d}})}]{footnote10}
\bibinfo{note}{In the wide-band limit a featureless (constant) electronic
  density of states of the reservoirs is assumed which has a bandwidth much
  larger than the characteristic energy scales of the system coupled to the
  reservoir. The retarded Green's function is then given by
  $G^R(\omega)=-\frac{1}{2W} \text{ln}\left(\frac{\omega-W}{\omega+W}\right)$
  where $W$ is the half bandwidth. To compare to the semi-circular DOS, the
  intra-lead hopping $t_\alpha^\text{wide-band}=\frac{t_\alpha}{4\pi}$ has to
  be rescaled so that the same $\Gamma$ is obtained.}

\bibitem[{\citenamefont{Knap et~al.}(2012)\citenamefont{Knap, Arrigoni, and
  von~der Linden}}]{knap_2012}
\bibinfo{author}{\bibfnamefont{M.}~\bibnamefont{Knap}},
  \bibinfo{author}{\bibfnamefont{E.}~\bibnamefont{Arrigoni}}, \bibnamefont{and}
  \bibinfo{author}{\bibfnamefont{W.}~\bibnamefont{von~der Linden}},
  \bibinfo{journal}{arXiv:1211.1384}  (\bibinfo{year}{2012}).

\bibitem[{\citenamefont{Sorantin et~al.}(2013)\citenamefont{Sorantin, Arrigoni,
  and von~der Linden}}]{sorantin2013}
\bibinfo{author}{\bibfnamefont{M.}~\bibnamefont{Sorantin}},
  \bibinfo{author}{\bibfnamefont{E.}~\bibnamefont{Arrigoni}}, \bibnamefont{and}
  \bibinfo{author}{\bibfnamefont{W.}~\bibnamefont{von~der Linden}}
  (\bibinfo{year}{2013}).

\bibitem[{\citenamefont{Jackson}(1998)}]{jackson}
\bibinfo{author}{\bibfnamefont{J.~D.} \bibnamefont{Jackson}},
  \emph{\bibinfo{title}{Classical Electrodynamics}} (\bibinfo{publisher}{Wiley
  New York}, \bibinfo{year}{1998}), ISBN \bibinfo{isbn}{978-0-47130-932-1}.

\bibitem[{\citenamefont{Schwinger}(1961)}]{schwinger_brownian_1961}
\bibinfo{author}{\bibfnamefont{J.}~\bibnamefont{Schwinger}},
  \bibinfo{journal}{J.Math. Phys.} \textbf{\bibinfo{volume}{2}},
  \bibinfo{pages}{407} (\bibinfo{year}{1961}).

\bibitem[{\citenamefont{Feynman and Jr.}(1963)}]{feynman_theory_1963}
\bibinfo{author}{\bibfnamefont{R.}~\bibnamefont{Feynman}} \bibnamefont{and}
  \bibinfo{author}{\bibfnamefont{F.~V.} \bibnamefont{Jr.}},
  \bibinfo{journal}{Ann. Phys.} \textbf{\bibinfo{volume}{24}},
  \bibinfo{pages}{118 } (\bibinfo{year}{1963}).

\bibitem[{\citenamefont{Keldysh}(1965)}]{keldysh_theory_1965}
\bibinfo{author}{\bibfnamefont{L.}~\bibnamefont{Keldysh}},
  \bibinfo{journal}{Zh. Eksp. Teor. Fiz.} \textbf{\bibinfo{volume}{47}},
  \bibinfo{pages}{1515} (\bibinfo{year}{1965}).

\bibitem[{\citenamefont{Rammer and Smith}(1986)}]{rammer_quantum_1986}
\bibinfo{author}{\bibfnamefont{J.}~\bibnamefont{Rammer}} \bibnamefont{and}
  \bibinfo{author}{\bibfnamefont{H.}~\bibnamefont{Smith}},
  \bibinfo{journal}{Rev. Mod. Phys.} \textbf{\bibinfo{volume}{58}},
  \bibinfo{pages}{323} (\bibinfo{year}{1986}).

\bibitem[{\citenamefont{Haug and Jauho}(1996)}]{haug_quantum_1996}
\bibinfo{author}{\bibfnamefont{H.}~\bibnamefont{Haug}} \bibnamefont{and}
  \bibinfo{author}{\bibfnamefont{A.}~\bibnamefont{Jauho}},
  \emph{\bibinfo{title}{Quantum Kinetics in Transport and Optics of
  Semiconductors}} (\bibinfo{publisher}{Springer-Verlag GmbH},
  \bibinfo{year}{1996}), \bibinfo{edition}{2nd} ed., ISBN
  \bibinfo{isbn}{3540616020}.

\bibitem[{foo({\natexlab{e}})}]{footnote3}
\bibinfo{note}{We denote real time by $\tau$.}

\bibitem[{\citenamefont{Gros and Valenti}(1993)}]{gros_cluster_1993}
\bibinfo{author}{\bibfnamefont{C.}~\bibnamefont{Gros}} \bibnamefont{and}
  \bibinfo{author}{\bibfnamefont{R.}~\bibnamefont{Valenti}},
  \bibinfo{journal}{Phys. Rev. B} \textbf{\bibinfo{volume}{48}},
  \bibinfo{pages}{418} (\bibinfo{year}{1993}).

\bibitem[{\citenamefont{S{\'e}n{\'e}chal
  et~al.}(2000)\citenamefont{S{\'e}n{\'e}chal, Perez, and
  Pioro-Ladri{\'e}re}}]{senechal_spectral_2000}
\bibinfo{author}{\bibnamefont{S{\'e}n{\'e}chal}},
  \bibinfo{author}{\bibfnamefont{D.}~\bibnamefont{Perez}}, \bibnamefont{and}
  \bibinfo{author}{\bibfnamefont{M.}~\bibnamefont{Pioro-Ladri{\'e}re}},
  \bibinfo{journal}{Phys. Rev. Lett.} \textbf{\bibinfo{volume}{84}},
  \bibinfo{pages}{522} (\bibinfo{year}{2000}).

\bibitem[{\citenamefont{Lanczos}(1951)}]{lanc.51}
\bibinfo{author}{\bibfnamefont{C.}~\bibnamefont{Lanczos}},
  \bibinfo{journal}{Journal of research of the National Bureau of Standards}
  \textbf{\bibinfo{volume}{45}}, \bibinfo{pages}{255} (\bibinfo{year}{1951}).

\bibitem[{\citenamefont{Bai et~al.}(1987)\citenamefont{Bai, Demmel, Dongarra,
  Ruhe, and van~der Vorst}}]{ba.de.87}
\bibinfo{author}{\bibfnamefont{Z.}~\bibnamefont{Bai}},
  \bibinfo{author}{\bibfnamefont{J.}~\bibnamefont{Demmel}},
  \bibinfo{author}{\bibfnamefont{J.}~\bibnamefont{Dongarra}},
  \bibinfo{author}{\bibfnamefont{A.}~\bibnamefont{Ruhe}}, \bibnamefont{and}
  \bibinfo{author}{\bibfnamefont{H.}~\bibnamefont{van~der Vorst}},
  \emph{\bibinfo{title}{Templates for the Solution of Algebraic Eigenvalue
  Problems: A Practical Guide (Software, Environments and Tools)}}
  (\bibinfo{publisher}{Society for Industrial and Applied Mathematics},
  \bibinfo{year}{1987}), ISBN \bibinfo{isbn}{0898714710}.

\bibitem[{\citenamefont{Economou}(2010)}]{economou_greens_2010}
\bibinfo{author}{\bibfnamefont{E.~N.} \bibnamefont{Economou}},
  \emph{\bibinfo{title}{Green's Functions in Quantum Physics}}
  (\bibinfo{publisher}{Springer}, \bibinfo{year}{2010}), \bibinfo{edition}{3rd}
  ed., ISBN \bibinfo{isbn}{3642066917}.

\bibitem[{\citenamefont{Nuss et~al.}(2012{\natexlab{b}})\citenamefont{Nuss,
  Arrigoni, Aichhorn, and von~der Linden}}]{PhysRevB.85.235107}
\bibinfo{author}{\bibfnamefont{M.}~\bibnamefont{Nuss}},
  \bibinfo{author}{\bibfnamefont{E.}~\bibnamefont{Arrigoni}},
  \bibinfo{author}{\bibfnamefont{M.}~\bibnamefont{Aichhorn}}, \bibnamefont{and}
  \bibinfo{author}{\bibfnamefont{W.}~\bibnamefont{von~der Linden}},
  \bibinfo{journal}{Phys. Rev. B} \textbf{\bibinfo{volume}{85}},
  \bibinfo{pages}{235107} (\bibinfo{year}{2012}{\natexlab{b}}).

\bibitem[{\citenamefont{Negele and Orland}(1998)}]{negele_quantum_1998}
\bibinfo{author}{\bibfnamefont{J.~W.} \bibnamefont{Negele}} \bibnamefont{and}
  \bibinfo{author}{\bibfnamefont{H.}~\bibnamefont{Orland}},
  \emph{\bibinfo{title}{Quantum Many-particle Systems}}
  (\bibinfo{publisher}{Westview Press}, \bibinfo{year}{1998}), ISBN
  \bibinfo{isbn}{0738200522}.

\bibitem[{foo({\natexlab{f}})}]{footnote4}
\bibinfo{note}{We use a the double adaptive integration scheme
  CQUAD.~\cite{Gonnet:2010:IRA:1824801.1824804}}.

\bibitem[{\citenamefont{Landauer}(1957)}]{Landauer_1957}
\bibinfo{author}{\bibfnamefont{R.}~\bibnamefont{Landauer}},
  \bibinfo{journal}{IBM Journal of Research and Development}
  \textbf{\bibinfo{volume}{1}}, \bibinfo{pages}{223} (\bibinfo{year}{1957}).

\bibitem[{foo({\natexlab{g}})}]{footnote8}
\bibinfo{note}{This effect has been obtained as well in
  \tcite{doi:10.1021/jz200862r}.}

\bibitem[{\citenamefont{Nakanishi and Tsukada}(1998)}]{Nakanishi_1998}
\bibinfo{author}{\bibfnamefont{S.}~\bibnamefont{Nakanishi}} \bibnamefont{and}
  \bibinfo{author}{\bibfnamefont{M.}~\bibnamefont{Tsukada}},
  \bibinfo{journal}{Jpn. J. Appl. Phys.} \textbf{\bibinfo{volume}{37}},
  \bibinfo{pages}{L1400} (\bibinfo{year}{1998}).

\bibitem[{\citenamefont{Nakanishi and Tsukada}(2001)}]{PhysRevLett.87.126801}
\bibinfo{author}{\bibfnamefont{S.}~\bibnamefont{Nakanishi}} \bibnamefont{and}
  \bibinfo{author}{\bibfnamefont{M.}~\bibnamefont{Tsukada}},
  \bibinfo{journal}{Phys. Rev. Lett.} \textbf{\bibinfo{volume}{87}},
  \bibinfo{pages}{126801} (\bibinfo{year}{2001}).

\bibitem[{\citenamefont{Ernzerhof et~al.}(2006)\citenamefont{Ernzerhof,
  Bahmann, Goyer, Zhuang, and Rocheleau}}]{doi:10.1021/ct600087c}
\bibinfo{author}{\bibfnamefont{M.}~\bibnamefont{Ernzerhof}},
  \bibinfo{author}{\bibfnamefont{H.}~\bibnamefont{Bahmann}},
  \bibinfo{author}{\bibfnamefont{F.}~\bibnamefont{Goyer}},
  \bibinfo{author}{\bibfnamefont{M.}~\bibnamefont{Zhuang}}, \bibnamefont{and}
  \bibinfo{author}{\bibfnamefont{P.}~\bibnamefont{Rocheleau}},
  \bibinfo{journal}{Journal of Chemical Theory and Computation}
  \textbf{\bibinfo{volume}{2}}, \bibinfo{pages}{1291} (\bibinfo{year}{2006}).

\bibitem[{foo({\natexlab{h}})}]{footnote5}
\bibinfo{note}{The large $U$ limit is directly accessible by Bethe
  Ansatz~\cite{PhysRevB.45.11795} or via transformation to tJ or Heisenberg
  Hamiltonians.}

\bibitem[{foo({\natexlab{i}})}]{footnote9}
\bibinfo{note}{Strictly speaking the threshold voltages $V_T$ depend on the
  Zeemann splitting which is however a very small energy scale for reasonable
  magnetic fields.}

\bibitem[{foo({\natexlab{j}})}]{footnote6}
\bibinfo{note}{In this work we refer to the single-particle gap $\Delta$ as the
  HOMO-LUMO gap also for interacting systems $U,W\neq0$.}

\bibitem[{foo({\natexlab{k}})}]{footnote7}
\bibinfo{note}{The corresponding field theory is exactly solvable, with eigen
  energies $\epsilon_n=\frac{\hbar^2 n^2}{2 m_e R^2}+V + \mu_B\sigma B
  -\frac{\hbar e}{2m_e}B n+\frac{e^2 R^2}{8 m_e}B^2$, where $V$ is the
  electrostatic potential and $R$ the radius of the ring. The (angular-)
  momentum $\hbar n$ is characterized by $n=0,\pm1,\pm2,\ldots$. The circular
  current driven by a magnetic field is given by $j_c=\frac{eN}{2\pi R
  m_e}(eB-\frac{2\hbar}{R^2}\sum\limits_{n_{\text{occ},\sigma}}n_\sigma)$.}

\bibitem[{\citenamefont{Schmitteckert}(2013)}]{Schmitteckert2013}
\bibinfo{author}{\bibfnamefont{P.}~\bibnamefont{Schmitteckert}},
  \bibinfo{journal}{Phys. Chem. Chem. Phys.} \textbf{\bibinfo{volume}{15}},
  \bibinfo{pages}{15845} (\bibinfo{year}{2013}).

\bibitem[{\citenamefont{Sch\"uler et~al.}(2013)\citenamefont{Sch\"uler,
  R\"osner, Wehling, Lichtenstein, and Katsnelson}}]{Schuler2013}
\bibinfo{author}{\bibfnamefont{M.}~\bibnamefont{Sch\"uler}},
  \bibinfo{author}{\bibfnamefont{M.}~\bibnamefont{R\"osner}},
  \bibinfo{author}{\bibfnamefont{T.~O.} \bibnamefont{Wehling}},
  \bibinfo{author}{\bibfnamefont{A.~I.} \bibnamefont{Lichtenstein}},
  \bibnamefont{and} \bibinfo{author}{\bibfnamefont{M.~I.}
  \bibnamefont{Katsnelson}}, \bibinfo{journal}{Phys. Rev. Lett.}
  \textbf{\bibinfo{volume}{111}}, \bibinfo{pages}{036601}
  (\bibinfo{year}{2013}).

\bibitem[{\citenamefont{Gonnet}(2010)}]{Gonnet:2010:IRA:1824801.1824804}
\bibinfo{author}{\bibfnamefont{P.}~\bibnamefont{Gonnet}}, \bibinfo{journal}{ACM
  Trans. Math. Softw.} \textbf{\bibinfo{volume}{37}}, \bibinfo{pages}{26:1}
  (\bibinfo{year}{2010}).

\bibitem[{\citenamefont{Yu and Fowler}(1992)}]{PhysRevB.45.11795}
\bibinfo{author}{\bibfnamefont{N.}~\bibnamefont{Yu}} \bibnamefont{and}
  \bibinfo{author}{\bibfnamefont{M.}~\bibnamefont{Fowler}},
  \bibinfo{journal}{Phys. Rev. B} \textbf{\bibinfo{volume}{45}},
  \bibinfo{pages}{11795} (\bibinfo{year}{1992}).

\end{thebibliography}

\end{document}